%% file: main.tex
\documentclass[acmlarge]{acmart}

\usepackage[english]{babel}

\usepackage{amsmath}
\usepackage{graphicx}

\usepackage{tikz, pgfplots}
\pgfplotsset{compat=1.18}
\usetikzlibrary {positioning}
\usepackage{forest}

\usepackage{subcaption}
\usepackage{listings}
\usepackage{cleveref}

\usepackage{tablefootnote}
\usepackage{hyperref}

\usepackage{graphicx}
\usepackage{subcaption}
\usepackage{listings}

\usepackage{array} 

\usepackage{xcolor}
\definecolor{sameeran}{RGB}{0, 150, 0}

\hypersetup{colorlinks=true, allcolors=blue}

\usepackage{tablefootnote}
\usepackage{hyperref}
\usepackage{tikz}

\usepackage{color, colortbl}

\definecolor{Gray}{gray}{0.9}

\input{textResources/lang-listing}

\newcommand\kw[1]{{\textbf{#1}}}

\title{Scheduling Languages: A Past, Present, and Future Taxonomy}

\author{Mary Hall}
\email{mhall@cs.utah.edu}
\affiliation{%
  \institution{University of Utah}
  \city{Salt Lake City}
  \country{USA}
}

\author{Cosmin Oancea}
\email{cosmin.oancea@di.ku.dk}
\affiliation{%
  \institution{University of Copenhagen}
  \city{Copenhagen}
  \country{Denmark}
}

\author{Anne C. Elster}
\email{elster@ntnu.no}
\affiliation{%
  \institution{Norwegian Institute of Science and Technology}
  \city{Trondheim}
  \country{Norway}
}

\author{Ari Rasch}
\email{a.rasch@uni-muenster.de}
\affiliation{%
  \institution{University of Muenster}
  \city{Muenster}
  \country{Germany}
}

\author{Sameeran Joshi}
\email{sameeran@cs.utah.edu}
\affiliation{%
  \institution{University of Utah}
  \city{Salt Lake City}
  \country{USA}
}

\author{Amir Mohammad Tavakkoli}
\email{amir.tavakkoli@utah.edu}
\affiliation{%
  \institution{University of Utah}
  \city{Salt Lake City}
  \country{USA}
}

\author{Richard Schulze}
\email{r.schulze@uni-muenster.de}
\affiliation{%
  \institution{University of Muenster}
  \city{Muenster}
  \country{Germany}
}
\begin{document}

\begin{abstract}
Scheduling languages express to a compiler a sequence of optimizations to apply.  Compilers that support a scheduling language interface allow exploration of compiler optimizations, \textit{i.e., exploratory compilers}.  While scheduling languages have become a common feature of tools for expert users, 
the proliferation of these languages 
without unifying common features may be confusing to users.  Moreover, we recognize a need to organize the compiler developer community around common exploratory compiler infrastructure, and future advances to address, for example, data layout and data movement.
To support a broader set of users may require raising the level of abstraction. This paper provides a taxonomy of scheduling languages, first discussing their origins in iterative compilation and autotuning, noting the common features and how they are used in existing frameworks, and then calling for changes to increase their utility and portability.  
\end{abstract}

\maketitle

\section{Introduction}

\enlargethispage{\baselineskip}

The main aim and contribution of this paper
is to provide a taxonomy of scheduling languages that illustrates how past work motivated and led to the present proliferation of scheduling languages, and how future improvements --- 
aimed at easing the interaction with the domain expert and at supporting more general forms of computations --- may cycle back to the past, prompted by the need for higher automation and integration. 
Table~\ref{tab:summary} summarizes the key properties of this evolution, which is highlighted in this 
section and expanded in the rest of the paper. 


Until the late 1990s, compilers were essentially black boxes that were controlled via optimization 
flags and a small set of directives to analysis and optimization.  Even today, large open source community compiler projects like LLVM and gcc are organized as a series of passes over a common intermediate representation (IR) that has been lowered from source code, with each pass leaving the code in a consistent state.  Therefore, the internals of each pass -- including the decision algorithms that apply optimizations -- are opaque to compiler users and even most compiler developers.  At the end of the compilation process, code is lowered from the IR to architecture-specific machine code.  

In the 1990s as compiler research introduced optimizations to achieve locality in caches (and vector and thread-level parallelism), 
something that is still true today became obvious: \textit{it is difficult to predict the best sequence of code transformations to achieve high performance since it depends heavily on both architecture and input data.}  
Many sophisticated cache models were developed during this time to predict capacity and conflict misses (e.g.,  interference phenomena~\cite{Temam:93} and cache miss equations~\cite{Ghosh99}) to guide architecture-specific optimization.    
However, 
as the complexity of these models grew,  
the alternative idea of simply executing the code to measure its performance on each platform, and adjust the algorithms accordingly, took hold. 
\textit{Autotuning}, popularized by ATLAS~\cite{ATLAS1,ATLAS2}, 
(Automatically Tuned Linear Algebra) ~\cite{ATLAS1,ATLAS2}, which focused on BLAS (Basic Linear Algebra) routines and a few LAPACK routines, initially focused on cache and instruction-level parallelism optimizations. Other 
similar approaches arose in the late 1990s and early 2000s for autotuning computation kernels from specific domains: for linear algebra~\cite{PhiPAC}, sparse linear algebra~\cite{Vuduc:PhD,SPARSITY}, Bitreversal 
~\cite{strandh-elster98, elster-meyer-bitrev09}, and FFTs~\cite{FFTW,SPIRAL1,SPIRAL2}.

This work on autotuning libraries was complemented by 
programming languages and compilers designed to facilitate exploration of more general computations. 
Contemporaneously with ATLAS, early work in \textit{iterative compilation} developed compiler technology that enabled exploration of code transformations~\cite{bodin1998iterative,Bodin2000IterativeCI,kisuki00,triantafyllis2003compiler,Pouchet08,Park11}.  Using iterative compilation, the compiler's internal algorithms were designed to explore alternative sequences of transformations, execute the code on the target hardware, and use that empirical data to decide which version provided the best performance.  Iterative compilation led to the reorganization of compilers to support exploration.  The search over transformation sequences was part of the compiler's algorithms, so it was still the compilers' responsibility to decide \textit{what to explore}.

\begin{table}
\caption{Motivations and resulting technology related to scheduling languages in the past, present, and expected future.}\vspace{-2ex}
\begin{tabular}{|p{0.65in}|p{1.4in}|p{1.4in}|p{1.8in}|} \hline
\textbf{Timeline} & \textbf{Motivation} & \textbf{Focus} & \textbf{Approaches}
\\ \hline
  \vbox{ \begin{center} \textsc{PAST} \bigskip\newline 1997 - 2012 \bigskip\newline \textsl{\textsc{Explore}} \end{center} }
  & 
  \vbox{ \begin{center} {\sc Heuristic-based code optimization decisions ineffective} \medskip\newline {\sc Improve efficiency of expert users} \end{center}}
  & 
  \vbox{ \begin{center} {\sc Primarily loop nest computations} \bigskip\newline {\sc Embedded} \& {\sc scientific applications} \& {\sc libraries} \end{center} }
  &  
  \vbox{ \begin{center} {\sc Autotuning libraries} \smallskip\newline {\sc Exploratory compilers and code generators} \smallskip\newline {\sc Rewritting rules} \&  {\sc lang. support for code variants} \end{center} }
\\[-9pt] \hline
  \vbox{ \begin{center}\textsc{PRESENT} \bigskip\newline 2013 - 2023 \bigskip\newline \textsl{\textsc{Specialize}} \end{center} }
  & 
  \vbox{ \begin{center} {\sc Efficiency of expert users in specific domains}\bigskip \end{center}}
  & 
  \vbox{ \begin{center} {\sc Domain specific languages and compilers}\bigskip \end{center} }
  &  
  \vbox{ \begin{center} {\sc Separate high-level specification and schedule} \medskip\newline {\sc Narrow the search space to utilize autotuning and ML} \end{center} }
\\[-9pt] \hline
  \vbox{ \begin{center}\textsc{FUTURE} \bigskip\newline 2024 -  \bigskip\newline \textsl{\textsc{Popularize}} \end{center} \vspace{-2ex}}
  & 
  \vbox{ \begin{center} {\sc Increase user accessibility} \end{center} \bigskip \begin{center} {\sc Raise abstraction} \bigskip\end{center} \vspace{-2ex}}
  & 
  \vbox{ \begin{center} {\sc Broaden to more general applications} \medskip\newline {\sc Unify and incorporate into common infrastructures}\end{center} \vspace{-2ex}}
  &  
  \vbox{ \begin{center} {\sc Data layout/movement integration} \medskip\newline {\sc Runtime support} \medskip\newline {\sc Expand search space while maintaining practicality}\end{center}\vspace{-2ex}} 
\\\hline
\end{tabular}
\label{tab:summary}
\end{table}

Subsequent research empowered expert programmers to collaborate with programming languages and compilers by exposing the optimization process to user control.   
On the programming language side, a number of systems allowed 
programmers to express alternative \textit{code variants} -- functionally equivalent implementations, potentially using different algorithms -- and to use autotuning to identify the most performant composition of variants
~\cite{sequoia,PetaBricks}.
On the compiler and code generation front, a number of systems supported the specification by an expert programmer of a sequence of transformations to apply to a code, expressed within the code using annotations or in a separate transformation recipe~\cite{Xlang,Girbal06,Chen2007CHiLLA,Yi07,Hartono09}.  

In the context of domain-specific languages (DSL),  Halide~\cite{10.1145/2491956.2462176} popularized the idea of deriving a high-performance implementation by combining: (1) a simple, high-level specification, aimed to increase productivity of domain experts; together with, (2) a transformation recipe written by a compiler expert in a separate language --- named a {\em scheduling} language.

Unlike standard interfaces to compiler optimizations -- compile-time flags and pragmas inserted in the code -- schedules are \textit{programs} that express a sequence of optimizations to apply to a separate source code.  Compilers that support a scheduling language interface allow exploration of compiler optimizations, and are referred to subsequently as \textit{exploratory compilers}. A survey of autotuning compiler approaches going back to the 1990s discusses exploratory compilers in more detail~\cite{AutotuningSurvey}.

The scheduling language serves as an expert programmer's interface to the compiler's transformations, or as an abstraction that exposes transformations to automated search and prediction.  Over the last decade, 
scheduling languages have been increasingly used to map performance-critical computations within domain-specific commercial applications. 
In principle, the use of a language to represent a schedule strengthens the optimization description by providing well-defined semantics, composability, and the opportunity for verification.  In practice, today's scheduling languages lack many of these~desirable~properties~of~a~programming~language~\cite{10.1145/3408974}.

In this paper, we consider the suitability of scheduling languages as a key abstraction in current and future compilers for achieving high performance on critical computations.  While scheduling languages were originally designed to expose optimizations to experts because automation was too difficult, we envision a future where scheduling languages are unified and part of any architecture-specific optimization workflow.  This exploratory compiler structure exposes profitable code transformation sequences at a finer granularity than tuning based on compiler flags, and is externally controllable in contrast to iterative compilation.  Moreover, a human-readable language facilitates improving and sharing schedules.   
Therefore, in the two decades since their origins, through the use of scheduling languages coupled with systematic approaches for exploring schedules, compilers have been restructured to provide the abstractions needed to close the circle and automate high-performance code generation.

The structure of this paper mirrors Table~\ref{tab:summary}. Section~\ref{sec:past} surveys past approaches ($1997-2012$) that were essentially aimed at optimizing scientific applications, beyond what the heuristic pipelines of general-purpose compilers could provide. They include autotuning libraries and advanced compilation techniques, such as iterative compilation, which are applied to mainstream languages and are characterized by a high degree of automation and integration. Section~\ref{sec:present} surveys present work ($2013$-$2023$) using Halide as the starting point. These systems are predominantly domain specific, driven by the observation that {\em specialization} of the language and compiler repertoire gives rise to performance. 
Section~\ref{sec:future} envisions a future that aims to broaden the specification language and the repertoire of code transformations available to scheduling, while at the same time interacting directly with the domain expert in simple(r) ways.  The future likely requires a high degree of automation, reminiscent of the approaches of the past. Section~\ref{sec:relwork} surveys related work, and Section~\ref{sec:conclusions} concludes.

\section{Past: Optimization Exploration}
\label{sec:past}
Early exploratory optimization systems arose from the growing complexity of emerging architectures and the productivity challenge of producing architecture-specific code.  These trends demanded that even expert programmers needed tools to accelerate exploration of different implementation strategies.  This section reviews the origins of scheduling languages by discussing three distinct bodies of work: domain-specific autotuning libraries and code generators; iterative compilation and early autotuning compilers and code generators; and programming languages that permitted exploration of optimizations and algorithms. 

\subsection{Early Domain-Specific Autotuning Library Generators}

Autotuning libraries are examples of early efforts to improve performance automatically (without programmer interactions) across different architectures.
These approaches 
decompose the given functions into performant subprograms, e.g., customized to 
the cache hierarchy of a given architecture. The sizes of the subprograms are parameterized and the parameters that offer best performance are determined empirically; thus,  \textit{autotuning} for a given architecture.
Some of the earliest efforts include PhiPAC,~ATLAS~and~FFTW. 

PhiPAC~\cite{PhiPAC} was a multi-level cache-blocked matrix multiply autotuning generator that was a precursor to ATLAS.
ATLAS (Automatically Tuned Linear Algebra) ~\cite{ATLAS1,ATLAS2}, focused on BLAS (Basic Linear Algebra) routines such as matrix-matrix multiplication and a few LAPACK routines. 
Autotuned compilation approaches combined with a small hand tuned assembly kernel, have also been shown to beat both ATLAS and vendor-optimized libaries ~\cite{jensen-karlin-elster2011}.
Since BLAS are performance-critical building blocks, the main vendors now offer BLAS implementations that are not only autotuned to their platforms, but also employ empirical based planners and optimizers, e.g. Intel´s MKL (Math Kernal Library) and Nvidia´s CuBLAS. 

Other efforts developed domain/algorithm specific libraries: such as for bitreversal~\cite{strandh-elster98, elster-meyer-bitrev09}, sparse linear algebra~\cite{Vuduc:PhD,SPARSITY} and FFTs. 
For example, the Fastest Fourier Transform in the West (FFTW) library~\cite{FFTW} takes advantage of the recursive nature of the FFT algorithm where smaller FFTs can be used as building blocks for larger FFTs. FFTW thus uses a high-level description execution plan for decomposing larger Fourier transforms into smaller,
specialized kernels named “codelets”, which can be tailored to the cache hierarchy. It then uses a dynamic programming-based search process at runtime, when the input transform size is known, to find the best execution plan.  Similiar techniques are used in CuBLAS. The Spiral~\cite{SPIRAL1,SPIRAL2} system applies to more general signal processing with high-level tensor notations and genetic algorithms-based~search.

\subsection{Exploratory Compiler and Code Generation Technology}

\subsubsection{Iterative Compilation}
Concurrent with the emergence of domain-specific autotuning 
libraries, \textit{iterative compilation} was developed as part of the OCEANS compiler project~\cite{OCEANS}.  Early work from this project by Bodin et al. describes the use of an iterative algorithm~\cite{bodin1998iterative,Bodin2000IterativeCI} that applied tiling, unrolling and padding to matrix multiply, and then searched among a fixed set of tile and unroll sizes. The portability of the search algorithm is demonstrated on three target architectures 
(Ultrasparc, Pentium Pro, and embedded VLIW Trimedia TM1000).
At that time, the high search cost of iterative compilation limited its use to embedded applications, where the assumption was that the application would be compiled once or infrequently and run repeatedly to amortize the search cost.  Kisuki et al. present a variety of search space algorithms and limits on searching tile and unroll sizes to make it practical for general-purpose optimization~\cite{kisuki00}.  
The \textit{optimization-space exploration (OSE)} compiler focused on VLIW optimizations for the Itanium processor, including VLIW-specific optimizations, standard loop transformations, and compiler flags~\cite{triantafyllis2003compiler}.  Pouchet et al. developed an iterative compilation approach that explored different valid multi-dimensional schedules using the polyhedral model, which facilitates correct composition of a sequence of iteration space transformations~\cite{Pouchet08}. Park et al. developed a prediction modeling technique called \textit{tournament predictor} to discover optimization sequences that outperformed -Ofast and other predictors using the Open64 compiler~\cite{Park11}.
%

A defining feature of iterative compilation research was that the search was part of the compiler's algorithms and not intended to be accessible to application developers. Triantafyllis et al. refer to the configuration of the iterative optimization algorithm as happening at \textit{compiler construction time}~\cite{triantafyllis2003compiler}.  The mechanism in these works to configure the compiler was not described in detail, but presumably looked very different from a scheduling language.

\subsubsection{Autotuning Compilers and Code Generators}

A few years later, motivated by the high performance enabled by autotuning libraries and iterative compilation, compilers and code generators emerged that made it possible for expert users to control the sequence of transformations applied to a computation.  The precursors to today's scheduling languages were focused on more general-purpose computation, typically limited to loop nests, where parallelizing compiler technology could be applied.  

\begin{figure}
\begin{tabular}{ll}
\begin{minipage}{.65\textwidth}
\begin{lstlisting}[basicstyle=\sffamily\footnotesize]
// code, named loops, xforms
#pragma xlang name iloop
for (i=0; i<NB; i++)
  #pragma xlang name jloop
  for (j=0; j< NB; j++)
    #pragma xlang name kloop
    for (k=0; k<NB; k++) {
      c[i][j] += a[i][k]*b[k][j];
    }
#pragma xlang transform stripmine iloop NU NUloop
#pragma xlang transform stripmine jloop MU MUloop
#pragma xlang transform interchange jloop NUloop
#pragma xlang transform interchange kloop NUloop
#pragma xlang transform fullunroll NUloop
#pragma xlang transform fullunroll MUloop
#pragma xlang transform scalarize_in b in kloop
#pragma xlang transform scalarize_in a in kloop
#pragma xlang transform scalarize_in&out c in kloop
#pragma xlang transform lift kloop.stores after kloop
\end{lstlisting}
\begin{center} 
(a) Xlang
\end{center}
\end{minipage}

&
\begin{minipage}{.35\textwidth}
\begin{lstlisting}[basicstyle=\sffamily\footnotesize]
// code
DO J=1,N
  DO K=1,N
    DO I=1,N
      C(I,J)=C(I,J)+
          A(I,K)*B(K, J)

// CHiLL
// transformation recipe
permute([3,1,2])
tile(0,2,TJ)
tile(0,2,TI)
tile(0,5,TK)
datacopy(0,3,2,[1])
datacopy(0,4,3)
unroll(0,4,UI)
unroll(0,5,UJ)
\end{lstlisting}
\begin{center} 
(b) CHiLL 
\end{center}
\end{minipage}

\end{tabular}\vspace{-1ex}
\caption{Examples of expressing locality optimizations for matrix multiply in Xlang\cite{Xlang} and  CHILL\cite{Chen2007CHiLLA}.} \label{fig:expr1}
\end{figure}

\paragraph{Expressing Transformations as an Interface to Compilers.}
An initial question was how to express the optimizations to be applied.  Initially, this approach permitted user access to compiler algorithms, and the scheduling language was primarily used to perform transformations on loop nest computations.  

The approach taken in the X Language was to express a sequence of transformations as pragmas in the source code~\cite{Xlang}, as in Figure~\ref{fig:expr1}(a).    An advantage of using pragmas is that the source code and optimization strategy are self-contained in a single file.  
However, a disadvantage is that only a single optimization strategy is supported.  The X Language also made it possible to define new transformations in the compiler so as to add capability to the language.

CHiLL~\cite{Chen2007CHiLLA,CHiLL-scalable,CHiLL-LCPC} (in \ref{fig:expr1}(b)) and URUK~\cite{Girbal06} supported a separate script that provided a sequence of transformations, each designating the source code statement to which the transformation should be applied along with parameters to the compiler's optimization.
The separate script has the advantage of supporting different optimization strategies or even architectures.

Figure~\ref{fig:expr1} shows how transformations are expressed in the X language and in CHiLL for matrix multiply.  In the X language, the loops are named so that transformations can use the names to designate where transformations are applied.  As new loops are created, such as with the \texttt{stripmine} transformation, they too are named.  In CHiLL's separate script, the \texttt{permute} command reorders the loop nest so that the I loop is outermost and the K loop is innermost, resulting in the dependence on C being carried by the innermost loop.   
For all but \texttt{permute}, the first parameter of each transformation is the statement to which the transformation should be applied.  The matrix multiply has a single statement, so it is always 0.  The second parameter is the loop, followed by any parameters to the transformation, e.g., the tile size or unroll factor.  Note that after the first \texttt{tile} command, 
a tile controlling loop is added in the outermost position (loop level 1).  Thus, in the next \texttt{tile} command, loop level 2 refers to the I loop, which was previously in the outermost position. Both the X Language and the CHiLL versions shown here are tiling i and j loops and unrolling some or all of the inner loop tiles.  CHiLL additionally tiles the k loop, and performs a datacopy of tiles of a and b into buffers to eliminate conflict misses in cache.  

All three frameworks -- X Language, URUK, and CHiLL -- exposed interfaces designed for expert users with knowledge about compiler transformations and abstractions.

\paragraph{Expressing Transformations as an Interface to Code Generators.}
Around the same time, 
code generators that emit specific code combined with existing templates were developed to apply transformations to code, typically integrated with autotuning.  Figure~\ref{fig:expr2} shows early examples of this approach as applied to matrix multiply,  ORIO~\cite{Hartono09} and POET~\cite{Yi07}.  Both examples apply unroll-and-jam for the loop nest computation.
On the right side of Figure~\ref{fig:expr2}(a), the inputs to the autotuner are provided, which include a series of problem sizes and unroll factors.  
In both systems, it is possible to define new transformations by describing how they modify the code.  For example, in Figure~\ref{fig:expr2}(b) \texttt{mm\_block\_unroll} combines blocking (i.e., tiling) and unrolling.  

\begin{figure}[t]


\begin{tabular}{ll}
\begin{minipage}{.4\textwidth}
 \vspace{0pt}
\begin{footnotesize}
\begin{verbatim}
/*@ begin Loop (
transform UnrollJam(ufactor=Ui)
for (i=0; i<=M-1; i++)
  transform UnrollJam(ufactor=Uj)
  for (j=0; j<=N-1; j++)
    transform UnrollJam(ufactor=Uk)
    for (k=0; k<=O-1; k++)
      A[i][j] += B[i][k]*C[k][j];
) @*/
for (i=0; i<=M-1; i++)
 for (j=0; j<=N-1; j++)
  for (k=0; k<=O-1; k++)
    A[i][j] += B[i][k]*C[k][j];
/*@ end @*/   
\end{verbatim}
\end{footnotesize}
\end{minipage}
&
\begin{minipage}{.4\textwidth}
\begin{footnotesize}
\begin{verbatim}
def performance_params {
  param Ui[] = range(1,33);
  param Uj[] = range(1,33);
  param Uk[] = range(1,33);
  constraint reg_capacity = Ui*Uj+Ui*Uk+Uk*Uj<=32;
}
def input_params {
  param M[] = [10,50,100,500,1000];
  param N[] = [10,50,100,500,1000];
  param O[] = [10,50,100,500,1000];
  constraint square_matrices = (M==N) and (N==O);
}
\end{verbatim}
\end{footnotesize}
\end{minipage}
\end{tabular}\\
(a) ORIO\\[9pt]

\begin{footnotesize}
\begin{verbatim}
<define loopJ Loop#("j",0,"n",1)>
<define loopI Loop#("i",0,"m",1)>
<define loopK Loop#("k",0,"l",1)>
<define mmStmt "c[i+j*m]+= alpha*b[k+j*l] * a[i+m*k];"> 
<define nest1 Nest#(loopK,mmStmt)>
<define nest2 Nest#(loopI,nest1)>
<define nest3 Nest#(loopJ,nest2)>
<define mmHead "void dgemm(int m, int n, int l, double alpha, double *a, double *b, double *c)">
<define dgemm Function#(mmHead, "int i, j, k;", nest3)> 
<define mm_block_unroll (Unroll#(mm_block, InnermostLoop#(mm_block)))>
<output mm_block_unroll.c (Block.bsize=(16 16 8);Unroll.ur=8; mm_block_unroll)>
\end{verbatim}
\end{footnotesize}
(b) POET\\
\caption{Examples of expressing matrix multiply in ORIO~\cite{Hartono09} and POET~\cite{Yi07}.}
\label{fig:expr2}
\end{figure}

\subsection{Language-Compiler Codesign}
\label{subsec:codesign-past}

\begin{figure}
\begin{subfigure}{0.55\columnwidth}
\begin{lstlisting}[basicstyle=\sffamily\footnotesize, numbers=left]
void task matmul::inner( in    float A[M][P]
                       , in    float B[P][N]
                       , inout float C[M][N] ){
  // tunable parameters specify the 
  // size of subblocks of A, B, C
  tunable int U, X, V;
  
  // Partition matries into sets of blocks
  blkset Ablks = rchop(A, U, X);
  blkset Bblks = rchop(B, X, V);
  blkset Cblks = rchop(C, U, V);
 
  // Compute all blocks of C in parallel
  mappar(int i=0 to M/U, int j = 0 to N/V) {
    mapreduce (int k=0 to P/X) {
        matmul( Ablks[i][k]    // recursive
              , Bblks[k][j]    // invocation
              , Cblks[i][j] ); // on subblocks
    }
  }
}

void task matmul::leaf( in    float A[M][P]
                      , in    float B[P][N]
                      , inout float C[M][N] ){
  for (int i=0; i<M; i++)
    for (int j=0; j<N; j++)
      for (int k=0; k<P; k++)
        C[i][j] += A[i][k] * B[k][j];
}
\end{lstlisting}\vspace{-1ex}
\caption{Algorithmic Specification}
\label{fig:sequoia-spec}
\end{subfigure}
\begin{subfigure}{0.42\columnwidth}
\begin{lstlisting}[basicstyle=\sffamily\footnotesize, numbers=left]
instance {
  name    = matmul_cluster_inst
  task    = matmul
  variant = inner
  run_at  = cluster_level
  calls   = matmul_node_inst
  tunable U = 1024, X = 1024, V = 1024
  A distribution = 2D block-block
            (blocksize 1024x1024) ... }
instance {
  name    = matmul_node_inst
  task    = matmul
  variant = inner
  run_at  = node_level
  calls   = matmul_L2_inst
  tunable U = 128, X = 128, V = 128   }
instance {
  name    = matmul_L2_inst
  task    = matmul
  variant = inner
  run_at  = L2_cache_level
  calls   = matmul_L1_inst
  tunable U = 32, X = 32, V = 32
  subtask arg A = copy
  subtask arg B = copy                }
instance {
  name    = matmul_L1_inst
  task    = matmul
  variant = leaf
  run_at  = L1_cache_level            }
\end{lstlisting}\vspace{-1ex}
\caption{Instantiation to a Cluster Hardware.}
\label{fig:sequoia-schedule}
\end{subfigure}\vspace{-1ex}
\caption{Running Example taken from Sequoia paper~\cite{sequoia}: specification (left) and schedule (right).}
\label{fig:sequoia}
\end{figure}

Complementary to exploratory compilers, this section provides a brief overview of a related strand of research that refers to enhancing the language with more powerful constructs as a way of lifting the level of abstraction at which the compiler reasons.

\subsubsection{Decomposition and Algorithmic Variants}
\label{subsubsec:algvariants-past}
Sequoia~\cite{sequoia} was one of the first works that proposed a specification language in which task decomposition is programmed explicitly --- but generically in terms of array sizes --- and a scheduling language that specializes the task to the particularities of the hardware by mapping the user-defined decomposition at each level of the memory hierarchy. Figure~\ref{fig:sequoia} shows the specification and schedule of the running example of the original paper. The specification of a Sequoia task consists of:
\begin{itemize}
\item recursive definitions (variant \texttt{inner} at lines $1-21$) that apply blocking primitives to generically-sized, multi-dimensional arrays (lines $9-11$) together with recursive calls to sub-blocks from inside map-reduce constructs (lines $16-18$);
\item a base-case definition (\texttt{leaf} at lines $23-30$) that expresses a sequentially-efficient implementation;
\item algorithmic variants are in principle supported at both levels;
\item communication is only possible between parent and child tasks by means of call-by-value-result semantics.
\end{itemize}

While the schedule in Figure~\ref{fig:sequoia-schedule} essentially performs tiling at each level of the memory hierarchy, 
the work on Sequoia has brought forth (at least) two key ideas: 

\begin{itemize}
\item[(1)] A slight strengthening of the language --- e.g., mostly side-effect free specification, isolation of communication and delegating the task decomposition to the user --- may simplify compiler's reasoning
and may result in simple implementations that offer competitive performance with state-of-the-art HPC libraries.
%
%
Many of the present DSLs borrow similar ideas from the data flow (or functional) context.


\item[(2)] Scheduling each level of the memory hierarchy allows to make explicit data-layout optimizations --- e.g., lines $24-25$ in figure~\ref{fig:sequoia-schedule} specify that arguments \texttt{A} and \texttt{B} of \texttt{matmul::leaf} are remapped in contiguous storage so as to ensure stride-$1$ accesses (and lines $8-9$ specify the data-distribution policy).   This is even more relevant today, e.g., because~GPUs~offer~programmable~memories~and~specialized~execution~units.
\end{itemize} 

Finally, a significant body of work was aimed at composing algorithmic variants, notably PetaBricks~\cite{PetaBricks} and the already discussed work aimed at FFT.
For example, PetaBricks demonstrates that the Poisson solver, symmetric eigenproblem and sorting can be efficiently implemented each from three algorithmic variants.
%

\subsubsection{Rewrite-Rule Systems in Purely-Functional Languages}
\label{subsubsec:functional-past}

Purely-functional languages primarily aim to provide a programming
environment that allows the implementation to match as closely as
possible the algorithmic specification. The support for higher
abstraction comes at the cost of (high) runtime overhead, which
would be prohibitively expensive unless it is statically optimized.
However, the low-level loop optimizations developed in the
imperative context are not applicable here because, for example,
recurrences are expressed in terms of recursive functions (rather
than loops) and type abstraction requires aggressive boxing,
which results in heavy use of indirection (pointers).
Instead, one of the directions taken has been to exploit
the richer semantics of higher-order operators by making
their {\em algorithmic} properties available to the compiler
under the form of re-write rules.
%
The key difference between re-write rules and affine transformations is that the former can express identities that cannot be derived by reordering the statements of the original code pattern. 

An example of such rewrite is the rule~\cite{nesl:vect-models} that famously states that a segmented scan~\cite{scan-primitive} (prefix sum) with an arbitrary associative operator---i.e., scanning in parallel each subarray of an irregular array of arrays--can be rewritten as a scan with a lifted operator that is applied to the array of tuples obtained by zipping the flattened-data array with a flag array that encodes with $1$ the start of each subarray (and $0$ otherwise).

\begin{figure}
\begin{subfigure}{0.48\columnwidth}
\begin{lstlisting}[basicstyle=\sffamily\footnotesize, mathescape=true, numbers=left]
foldr :: ($\alpha$ -> $\beta$ -> $\beta$) -> $\beta$ -> [$\alpha$] -> $\beta$
foldr k z [] = z
foldr k z (x:xs) = f x (foldr f z xs)

build:: (forall b. (a -> b -> b) -> b -> b)
        -> [a]
build g = g (:) []

{-# RULES
"foldr/build"
forall (g:: $\forall$ b. (a -> b -> b) -> b -> b)
       k z.
foldr k z (build g) = g k z
#-}
\end{lstlisting}
\end{subfigure}
\begin{subfigure}{0.49\columnwidth}
\begin{lstlisting}[basicstyle=\sffamily\footnotesize, mathescape=true, numbers=left]
sum :: [Int] -> Int    -- sum [5,4,3,2,1] = 15
sum xs = foldr (+) 0 xs

down :: Int -> [Int]   -- down 5 = [5,4,3,2,1]
down v = build ($\lambda$ n -> down' v n)

down' 0 cons nil = nil
down' v cons nil = cons v (down' (v-1) cons nil)

sum (down 5)
$\Rightarrow$ -- inlining sum and down
  foldr (+) 0 (build (down' 5)) 
$\Rightarrow$ -- applying foldr/build re-write
  down' 5 (+) 0
  -- computes 5 + (4 + (3 + (2 + (1 + 0))))
\end{lstlisting}
\end{subfigure}\vspace{-2ex}
\caption{Example of Haskell Re-Write Rule from~\cite{plying-by-rewrites}: Shortcut deforestation rule (left) and its application (right).}
\label{fig:haskell-rules}
\end{figure}

A large body of work~\cite{optran,opal,tampir,stratego,mag} has
studied in the functional context how to allow a library writer
to extend the compiler by means of re-write rules that encode 
``domain-specific'' optimizations. 
The Haskell GHC compiler employs such a mechanism that has passed
the test of time and differentiates itself from related approaches
by means of its practical simplicity~\cite{plying-by-rewrites}. 
Re-write rules are written in source-to-source Haskell, 
that enables simple pattern matching and a simple
re-write strategy. 

An example of re-write rule (used in~\cite{plying-by-rewrites}) is shown at the bottom left of 
figure~\ref{fig:haskell-rules}: it consists of (1) a name 
{\tt ``foldr/build''}, (2) a {\tt forall} clause that
declares which of the variables used in the rule are universally
quantified, and (3) a body in which the left hand side
must be the application of a function which is not one of the
quantified variable ({\tt foldr} in our case).  When the rule
fires, the left hand side is always replaced with the right-hand
side; it is the user's responsibility to ensure that (1) the rule
is correct, (2) the right-hand side is more efficient than the
left-hand side, and that (3) the system of re-writes terminates.

The rule uses the higher-order functions {\tt foldr} and {\tt build}. {\tt foldr} is conventionally defined, i.e., it has the semantics
$\kw{foldr}~\odot~z~[x_1,\ldots,x_n] \ = \ x_1 ~\odot~ (x_2 ~\odot~ ... (x_n \odot z))$.
{\tt build} takes as parameter a functional
representation of a list {\tt g} --- that abstracts over its cons and nil constructors --- and it applies {\tt g} to ordinary
list constructors {\tt :} and {\tt []}. The rule states that the cases when {\tt foldr} consumes the application of {\tt build}  to {\tt g} as its third argument, can be re-written by applying
{\tt g} directly. The right hand side of figure~\ref{fig:haskell-rules}, 
demonstrates how this rule eliminates the creation of the
intermediate list $[1,2,3,4,5]$.  

Essentially, the GHC performs the boring transformations --- such
as beta reduction, inlining, case switching, let floating, case
swapping and elimination --- and relies heavily on the library 
writer to directly communicate the smarts to the compiler. 
Common examples include fusion rules for user-defined ADTs 
(e.g, rose trees) and specialization rules for (common cases of)
overloaded instances, where special attention is dedicated to 
eliminating the overheads of maintaining a modular programming style.
Notably, the compiler also automatically generates and then
applies re-write rules, typically pertaining to specializations.

\section{Present: What's in a Schedule?}
\label{sec:present}


Our taxonomy categorizes the present as beginning with Halide~\cite{ragan2013halide}, which coined the phrase ``separate schedules from the underlying algorithm'' and played an important role in popularizing scheduling languages.  The present, which represents a decade of progress, has coincided with the recognition that \textit{specialization gives rise to efficiency}, leading to a proliferation of domain-specific hardware and software tools.  Consequently, we organize this section into a set of key domains where scheduling languages were paramount to unlocking high performance. Of note, the systems described here increasingly target mainstream commercial applications, while past approaches of Section~\ref{sec:past} were developed in the context of scientific computing or embedded systems. 

\begin{figure}[t]
\includegraphics[width=\columnwidth]{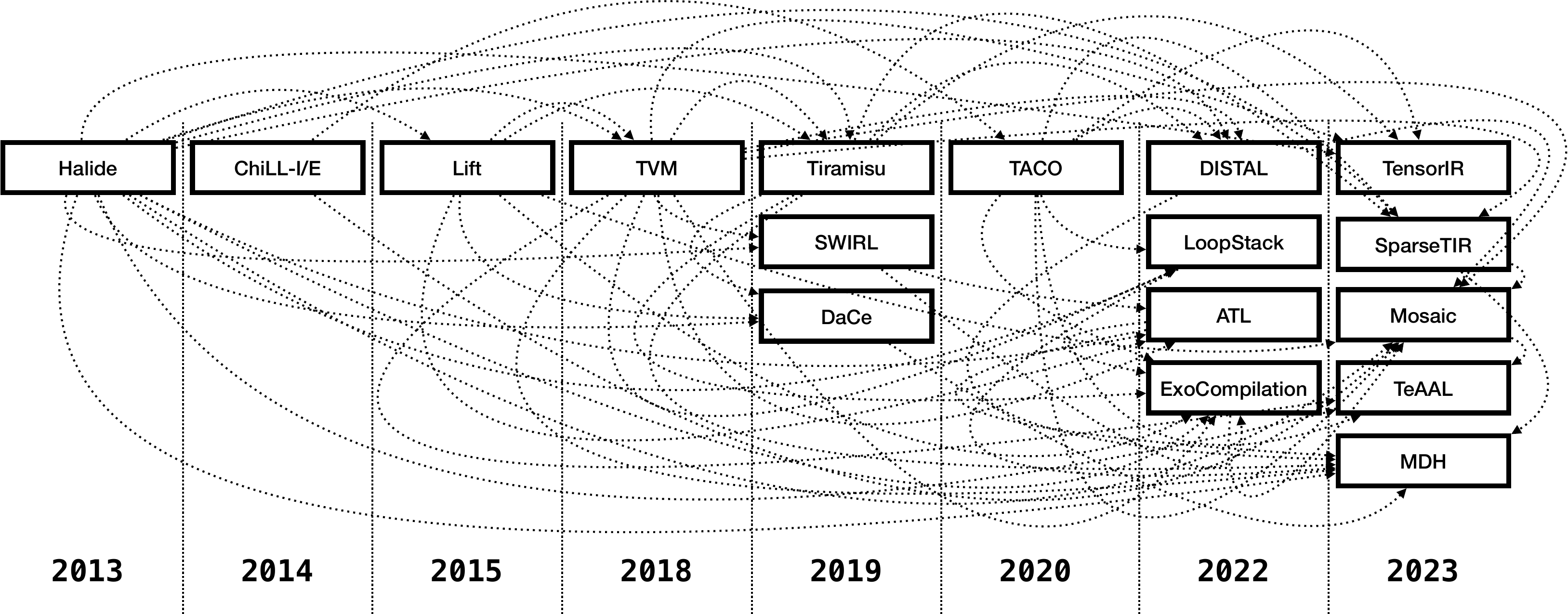}\vspace{-1ex}
\caption{
Evolution of a number of scheduling approaches (arrows indicate citation). 
}
\label{fig:timeline}
\end{figure}

The evolution of a number of scheduling approaches is depicted in Figure~\ref{fig:timeline}, which uses arrows to denote citations in the pointed-to paper and thus represents relationships in a visual manner. Each subsection discusses an application domain, complemented with figures 
that summarize how the discussed central works have influenced each other, i.e., the evolution that provided more context and exposed more optimizations.

\begin{figure}[H]
\begin{center}
\resizebox{\columnwidth}{0.7in}{
\begin{tikzpicture}[
SRP/.style={rectangle, draw=purple!60, fill=purple!5, very thick, minimum size=1cm},
SRR/.style={rectangle, draw=red!60, fill=red!5, very thick, minimum size=1cm},
]

\begin{small}
\node[SRP, text width=7cm]    (Halide)                              {\textbf{Halide (2013)} -- \linebreak
\textbf{Targets:} fusion of image processing pipelines
\linebreak
\textbf{Scheduling:} granularity of compute and store,
 \linebreak
split, vectorize, reorder, parallelize};

\node[SRP, text width=7cm]    (PolyMage)  [right=of Halide]
{\textbf{PolyMage(2015)} -- 
\linebreak \textbf{Targets:} same as Halide 
\linebreak
\textbf{Scheduling:} affine transformations (various tiling) + greedy grouping procedure
};
\end{small}

\draw[->, very thick] (Halide.east)  to node[right] {} (PolyMage.west);


\
\end{tikzpicture}}\vspace{-2ex}
\caption{Overview of two of the central works related to DSL scheduling languages targeting image processing.
}
\label{fig:sched-lang-overview-ip}
\end{center}
\end{figure}
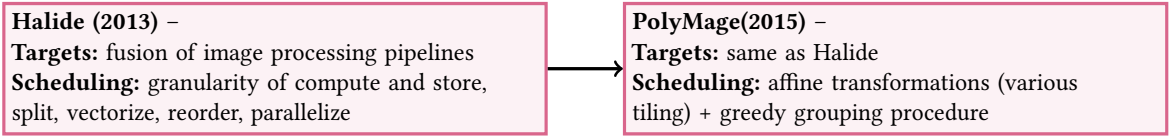

\subsection{Image Processing Approaches (summarized in \cref{fig:sched-lang-overview-ip})}
\textbf{Halide}~\cite{10.1145/2491956.2462176} is one of the first works to pragmatically combine the strengths of the functional and imperative approaches in the
context of a DSL that is primarily aimed at image-processing pipelines. 

%

%
The key idea in Halide is to separate the algorithm's expression from the optimization concerns by using:
\begin{itemize}
\item a simple and pure data-flow specification that is implicitly parallel and accessible to the domain expert, 
\item an optimization recipe that is written by the compiler expert or is derived by 
autotuning.
\end{itemize}

Figure~\ref{fig:halide-spec} shows the expression of a $3\times 3$
un-normalized box filter, in which arrays (e.g., {\tt blurx}, {\tt out})\footnote{
  {\tt blurx} and {\tt out} compute the horizontal and isotropic blur by averaging over $3\times 1$ and $1\times 3$ windows, respectively.
}
are represented as index functions (i.e., from coordinates to values), hence their elements can be safely computed in parallel.
In this context, the optimization of highest impact is {\em stencil fusion},
which can be realized by a combination of tiling, sliding-window and
work replication strategies.\footnote{
  These techniques cover a trade-off space along three axes:
  the degree of locality, of exploited parallelism and of redundant
  computation.
}
The best combination is however sensitive to the hardware and dataset
characteristics.
As such, the optimization recipe answers the questions: (1) at which
granularity to {\em compute} and (2) at which granularity to {\em store}
each of the intermediate arrays, and, within those grains, (3)
in what order/fashion should the array domain be traversed?\footnote{
  For example, dimensions can be strip-mined, reordered,
  traversed sequentially or in parallel (vectorization included).
}
For example, the optimization recipe in figure~\ref{fig:halide-opt-recipe}
declares a schedule that combines the tiling and sliding window optimizations:
\begin{itemize}
\item {\tt out} is tiled with tiles of size $4$, creating (re-ordered) dimensions of indices $y_o$, $x_o$, $y_i$, $x_i$, from which $y_o$, $x_o$ are parallelized, $y_i$ is sequentialized (to implement a sliding window) and $x_i$ is vectorized.
\item {\tt blurx} is {\em stored} at the level of loop $y_o$ in {\tt out} but is {\em computed} at a finer granularity ($y_i$) --- hence the domain of {\tt blurx} under $y_i$ only has dimension $x_i$, which is vectorized for performance.
\end{itemize}

\begin{figure}[h]
\begin{subfigure}{0.48\columnwidth}
\begin{lstlisting}[basicstyle=\sffamily\footnotesize, mathescape=true, numbers=left]
UniformImage in(Uint(8), 2)
Var x, y
Func blurx(x,y) = in(x-1,y)
                + in(x,y)
                + in(x+1,y)
Func out(x,y) = blurx(x,y-1)
              + blurx(x,y)
              + blurx(x,y+1)
\end{lstlisting}
\caption{Data-flow specification of a\\ $3\times 3$ un-normalized box filter }
\label{fig:halide-spec}
\end{subfigure}
\begin{subfigure}{0.48\columnwidth}
\begin{lstlisting}[basicstyle=\sffamily\footnotesize, mathescape=true, numbers=left]
blurx: split x by 4 $\rightarrow$ $x_o$, $x_i$
       vectorize: $x_i$
       store at out.$x_o$
       compute at out.$y_i$
out: split x by 4 $\rightarrow$ $x_o$, $x_i$
     split y by 4 $\rightarrow$ $y_o$, $y_i$
     reorder: $y_o$, $x_o$, $y_i$, $x_i$
     parallelize: $y_o$
     vectorize:   $x_i$
\end{lstlisting}
\caption{Optimization Recipe (i.e., Schedule Specification)}
\label{fig:halide-opt-recipe}
\end{subfigure}\vspace{-2ex}
\caption{Halide running example~\cite{10.1145/2491956.2462176}.}
\label{fig:halide}
\end{figure}

The legality of the transformations is enabled by the pure (and implicitly parallel) semantics of the array computations, e.g., tiling is legal because parallel loops are always safe to be interchanged inwards.

In summary, Halide has demonstrated that high-performance implementations of image-processing pipelines can be derived by means of a simple and clean DSL specification in conjunction with (1) either an optimization recipe written by the compiler expert or with (2) extensive autotuning. Initially, its autotunner used a genetic algorithm, but its slow convergence motivated the switch to the more robust OpenTuner framework~\cite{opentuner}; this, however, still required hours-to-days to find the optimal solution for deep pipelines.

\vspace{0.07in}   \noindent
\textbf{PolyMage}´s
approach ~\cite{PolyMage} was (partly) motivated by the observation that even though the schedule space is vast,\footnote{The schedule (search) space is exponential in the depth of the pipeline, i.e., the number of pipeline stages.} only a smallish subset of that space matters in practice. As such, PolyMage renounced the optimization recipe in favor of a {\em greedy grouping procedure}, which aggressively fuses computation until a maximal threshold of redundant computation is reached.
Since the grouping procedure is parameteric in terms of tile and threshold sizes, the autotuning is simplified to explore a smallish space of (three) threshold values and (seven) tile sizes per tiled dimension.
This procedure is reported to find in up to one hour a near-optimal schedule for multi-core execution that offers performance competitive to code written/optimized by experts.
Finally, PolyMage demonstrated that polyhedral reasoning can be applied in a DSL context to elegantly model 
\begin{itemize}
\item overlapped tiling --- which, in principle, falls outside polyhedral scheduling since it introduces redundant computation (i.e., does not preserve the statements of the original program), and
\item other classical tiling strategies --- such as parallelogram, hexagonal or split tiling -- which were not feasible to be expressed in Halide. 
\end{itemize}

Since then, a rich body of work has been  aimed at improving Halide in various ways, for example (1) by adapting the scheduling strategy of PolyMage to generate competitive Halide schedules~\cite{halide-tuning-in-secs}, (2) by extending Halide to generate code for Specialized Digital Signal Processors (DSPs) compilers~\cite{extending-halide-for-DSP}, (3) by providing support for automatic computation of gradients~\cite{halide-ad}, (4) by adding new optimization strategies to maximize producer-consumer locality~\cite{autogen-halide-schedules-to-max-loc-by-reuse}, (5) by supporting automatic generation of GPU schedules~\cite{Halide-gpu-sched}, and (6) by refining the search algorithm to quickly produce high-quality schedules~\cite{halide-tree-search-and-random-prgs,halide-search-scalability}. \textbf{ImageCL}~\cite{imagecl-HPCS-16, falch-elster-cpe.4384}
also showed improvements over Halide and others.

\begin{figure}[h]
\begin{center}
\resizebox{\columnwidth}{1.5in}{
\begin{tikzpicture}[
SRB/.style={rectangle, draw=blue!60, fill=blue!5, very thick, minimum size=1cm},
]

\begin{small}
\node[SRB, text width=7.7cm]    (TCE)           {
\textbf{TCE(2002)} -- \linebreak
\textbf{Targets:} dense tensors 
\linebreak
\textbf{Scheduling:} compiler exploration instead of scheduling
 };

\node[SRB, text width=7cm]    (Barracuda)  [right=of TCE]
{\textbf{Barracuda (2015)} -- 
\linebreak \textbf{Targets:} GPU + domain-specific optimizations
\linebreak         
\textbf{Scheduling:} CUDA-CHILL and Orio
};

\node[SRB, text width=7cm]    (Distal)       [below=of TCE] 
{\textbf{Distal (2022)} -- 
\linebreak  \textbf{Targets:} CPU + GPU clusters
\linebreak
\textbf{Scheduling:} loop transformation + 
specification of data distribution and communication
};
\end{small}
\draw[->, very thick] (TCE.east)  to node[right] {} (Barracuda.west);
\draw[->, very thick] (TCE.south)  to node[below] {} (Distal.north);
\end{tikzpicture}}
\medskip

\resizebox{\columnwidth}{1.7in}{
\begin{tikzpicture}[
SRG/.style={rectangle, draw=green!60, fill=green!5, very thick, minimum size=1cm},
]
\begin{small}
\node[SRG, text width=8cm]    (CHILL)         {\textbf{CHiLL-I/E (2014)} -- \linebreak
\textbf{Targets:} sparse tensors 
\linebreak
\textbf{Layouts:} CSR, ELL, DIA, BCSR \linebreak
 \textbf{Scheduling:} inspector/executor+sparse polyhedral 
 e.g. coalesced, parallel, reduction, make-dense, tile, skew
 }; 

\node[SRG, text width=7cm]    (TACO)  [right=of CHILL]
{\textbf{TACO (2017)} -- \linebreak 
\textbf{Targets:} co-iteration over multiple tensors
\linebreak  
 \textbf{Layouts:} per dimension sparse/dense,
 \linebreak
 COO, ELL, DIA, hashmap
\linebreak 
\textbf{Scheduling:} + pos, coord, precompute, bound
};

\node[SRG, text width=7cm]    (Mosaic)       [below =of CHILL] 
{\textbf{Mosaic (2023)} -- 
\linebreak  
\textbf{Targets:} integration of optimized library kernels
\linebreak
   \textbf{Scheduling:} Bind, Map
};

\node[SRG, text width=9cm]    (TeALL)       [right =of Mosaic] 
{\textbf{TeALL(2023)} -- \linebreak  
\textbf{Targets:} concise specification for accelerators
\linebreak
 \textbf{Scheduling:} rank-order, partitioning, loop-order, space-time
};
\end{small}
\draw[->, very thick] (CHILL.east)  to node[right] {} (TACO.west);
\draw[->, very thick] (TACO.south)  to node[right] {} (Mosaic.north);
\draw[->, very thick] (TACO.south)  to node[right] {} (TeALL.north);

\end{tikzpicture}}
\vspace{-4ex}
\caption{Overview of some central works related to scheduling languages for dense (blue) and sparse tensor algebra (green).
}
\label{fig:sched-lang-overview-tensor}
\end{center}
\end{figure}
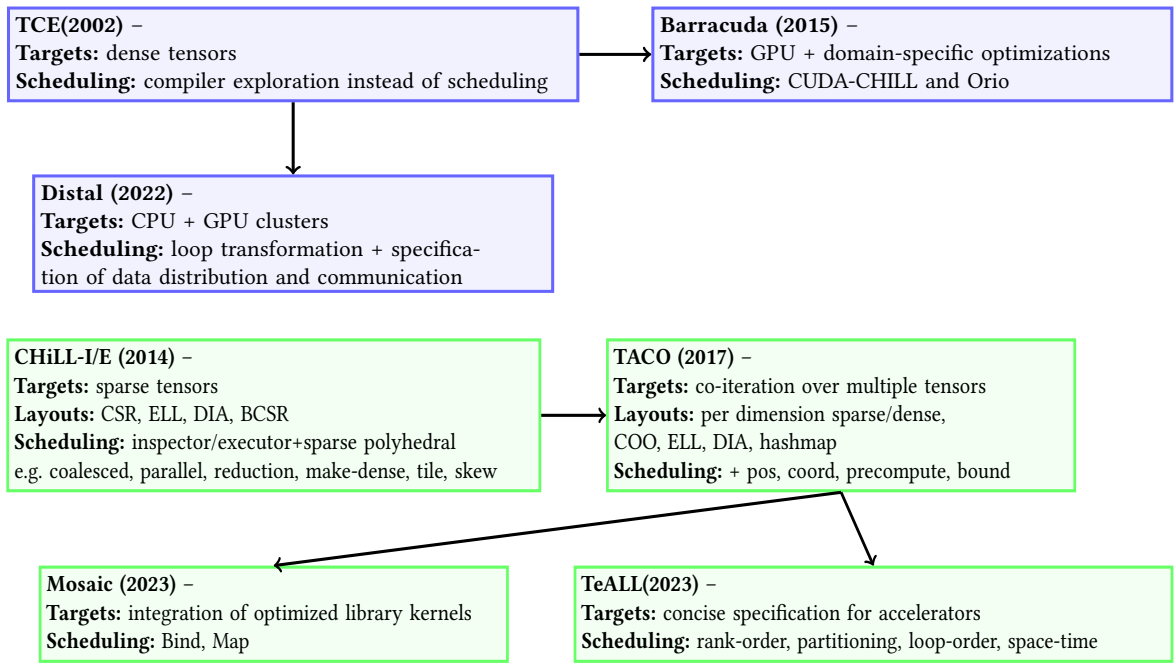

\subsection{Tensor Algebra Approaches (summarized in \cref{fig:sched-lang-overview-tensor})}

Tensor algebra describes an algebra applied to multi-dimensional tensors of any rank, with multiplication used to compute tensor products.  Tensor contraction refers to multiplying elements of two tensors to produce a third tensor; a contraction dimension results in a summation of the products of the two tensors along that dimension.  In this section, we separate 
dense and sparse tensor algebra, as the expression and optimization of them varies significantly.  While dense tensor algebra exhibits high arithmetic intensity and demands optimizations to manage the memory hierarchy and parallelism, sparse tensor algebra suffers from being memory bound and involves a significantly more complex code generation process.


\subsubsection*{Dense Tensor Algebra}
Beyond the various strategies for implementing matrix multiply, we focus this section on support for tensor contraction for higher-dimensional tensors.  An early domain-specific system for tensor algebra was the \textbf{Tensor Contraction Engine (TCE)}~\cite{TCE1,TCE3}, designed for a class of quantum chemistry computations.  These are characterized by contractions over 4-dimensional tensors with hundreds of terms.  The computations are both compute-intensive, and access a large volume of data.  The order in which the tensor contraction is performed impacts the overall amount of computation.  Moreover, with so many terms with different dimension orders, many address streams are touched during the computation.   These challenges are addressed using enumeration of feasible solutions that fit in memory, minimization of total computation, and fusion heuristics to limit the set of implementations explored~\cite{TCE1,TCE3}.  While work on TCE predates the use of scheduling languages in compilers, it employs search and search space pruning to explore a prohibitively large set of feasible implementations.

\textbf{Barracuda}~\cite{Nelson15} is a more recent work on 4-dimensional tensor contraction, optimizing quantum chemistry kernels and nuclear fusion computations on GPUs.  The approach uses exploratory compilation, from mathematical description to optimized CUDA code as output.  Starting with a high-level tensor input language, the approach combines tensor-specific mathematical transformations with a GPU decision algorithm, and autotuning of a large parameter space using a random forest search algorithm. Internally, this implementation uses CUDA-CHiLL's scheduling language, and Orio to manage the search algorithm.  A key heuristic to limit the search space is to choose loop orders that match memory layout order for at least one of the tensors in the computation, thus limiting data movement.  

\textbf{Fireiron}~\cite{hagedorn2020fireiron} introduces a scheduling language that is fully focused on optimizing dense matrix multiplication on NVIDIA GPUs.
For this, Fireiron introduces scheduling primitives to target domain-specific hardware extensions of NVIDIA GPUs, namely Tensor Cores which compute matrix multiplication of small $4\times4$ matrices immediately in hardware, thereby enabling high performance potentials.   
However, while Fireiron achieves impressive performance results for its particular target domain~-~multiplication of dense matrices on NVIDIA GPUs~--~it cannot be used for other computations and/or architectures, which limits its applicability.

\textbf{DISTAL} represents the state-of-the-art in scheduling languages for tensor algebra, targeting CPU and GPU clusters~\cite{DISTAL}.  The scheduling language in DISTAL includes loop transformations to organize the computation as well as communication specification across distributed nodes. 
Moreover, a separate format description permits data layout within a node and distributed across nodes.
This approach makes it possible to specify common tensor algebra algorithms such as Cannon and COSMA at a high level, and automate their generation to optimized code.

\subsubsection*{Sparse Tensor Algebra}
Sparse tensor algebra performs the same operations as dense tensor algebra, but the underlying data representation is fundamentally different.  When many entries of a tensor are zero, sparse tensor representations only store nonzero values, to (1) reduce the size of data; (2) avoid unnecessary computation such as multiplying by zero or adding zero; and, (3) reduce data movement through the memory hierarchy.  Auxiliary data structures provide a mapping of nonzero values to their logical indices in a dense matrix.  This physical-to-logical mapping makes it possible for sparse tensor code generators to perform the portions of a computation on the nonzero values.  

As compared to dense tensor algebra, we observe that scheduling languages for sparse tensor algebra have several unique capabilities.  First, they support a variety of sparse data representations -- hereafter referred to as \textit{data layouts} -- since proper code generation must be customized to a layout.  Moreover, loop optimizations must be reformulated whenever loop indices iterate over a sparse dimension of a tensor.  They may also require constructs that utilize information only available at runtime into the optimization.  These points will be illustrated by examples from sparse tensor algebra systems that employ scheduling languages.

\textbf{CHiLL-I/E} extended
CHiLL's transformations and associated scheduling commands to support sparse linear algebra computations~\cite{venkat:cgo14,venkat:pldi15,venkat:ipdps16,venkat:sc16,AhmadVH16}.  CHiLL was extended to incorporate concepts from the sparse polyhedral framework~\cite{STROUT201632}, such as uninterpreted functions representing index arrays.  CHiLL-I/E was able to convert sparse matrix representations using an \textit{inspector/executor paradigm}, whereby a one-time inspector at runtime converted the matrix from a standard representation to an optimized one, and the executor was then able to generate optimized GPU or CPU code.
The scheduling language constructs added to CHiLL included standard transformations such as \textit{coalesce} (also called collapse and flatten), \textit{parallel} and \textit{reduction}~\cite{venkat:cgo14}.
In addition, the 
\textit{make-dense} transformation designated a sparse dimension as being dense as a way for the compiler to reason about subsequent transformations. These included a sequence of standard transformations such as \textit{tile} and \textit{skew}.  Ultimately, any \textit{make-dense } was proceeded by  \textit{compact}, or \textit{compact-and-pad} transformations, which generate both an inspector for format conversion and an executor for parallel execution on a GPU. This work enabled conversion from CSR format to common formats ELL, DIA and BCSR~\cite{venkat:pldi15}. Uninterpreted functions were also used to incorporate dynamic parallel wavefronts~\cite{venkat:sc16}.

While CHiLL-I/E transformations were designed to support a single sparse tensor, \textbf{TACO}~\cite{kjolstad2017tensor} introduced an approach to co-iteration over multiple sparse tensors, where the intersection (for multiply) or the union (for addition) of the nonzero locations must be identified.  The user specifies the layout  along with the 
computation in Einstein notation, and the compiler generates the code for the input with the specified layout.  Originally, TACO represented data layouts by marking dimensions as being either \textit{dense} or \textit{compressed}, where compressed dimensions only represented nonzero values; a similar approach was integrated into the MLIR compiler~\cite{Bik:MLIR}. Subsequent extensions incorporated  \textit{singleton}, \textit{range}, \textit{offset}, and \textit{hash} to support other common sparse tensor layouts that are the higher-dimensional analogs of sparse matrix representations COO, ELL, and DIA, as well as a hashmap~\cite{chou2018format}. \cite{ahrens2023looplets} extends the layout specification to describe regular and irregular patterns. Code transformations were subsequently enabled in {TACO} through a scheduling language~\cite{senanayake2020sparse}. 
Key transformations are \textit{split} (i.e., strip-mine), and \textit{collapse} (also called coalesce or flatten), \textit{reorder} (i.e., permute), \textit{unroll} and \textit{parallelize}.
The transformations \textit{pos} and \textit{coord} enable transformations to be applied to the position (or physical) space, and the coordinate (or logical) space.  The transformation \textit{precompute} permits subarrays to be computed in scratchpad memories, and \textit{bound} introduces constants used by code generation.
The transformations are applied on an iteration graph IR before sparse code generation.


Two systems for sparse tensor computations extend the use of scheduling languages in unique ways.  
\textbf{Mosaic}\cite{bansal2023mosaic} 
extends {TACO}'s scheduling language to combine code generation and optimization with integration with optimized library kernels.  
For this purpose, Mosaic introduces two essential scheduling commands, \texttt{Bind}, which indicates that a statement should be replaced by a function call, and \texttt{Map}, which provides partial automation of schedule generation. 
The Tensor Algebra Accelerator Language (\textbf{TeAAL}) 
uses a scheduling language
to enable precise and concise specification of sparse tensor algebra accelerators, and inspiration comes from dense tensor algebra accelerator design\cite{nayak2023teaal}. 
Examples of scheduling language primitives include rank-order, partitioning, loop-order, and spacetime. 

\begin{figure}[h]
\begin{center}
\resizebox{\columnwidth}{1.8in}{
\begin{tikzpicture}[
SRY/.style={rectangle, draw=yellow!60, fill=yellow!5, very thick, minimum size=2cm}
]
\begin{small}
\node[SRY, text width=7cm]    (Lift)      
{\textbf{Lift (2015)} 
: purely-functional, map-reduce\linebreak  
 \textbf{Targets:} mainly GPU hardware 
\linebreak
\textbf{Scheduling:} based on functional rewrite rules
};

\node[SRY, text width=7cm]    (Tiramisu)   [right =of Lift]   
{\textbf{Tiramisu (2019)}
: C++ embedded DSL\linebreak
\textbf{Targets:} multiple platforms including multicore, GPU, FPGA,  distributed systems
\linebreak
\textbf{Scheduling:} polyhedral transformations 
};

\node[SRY, text width=7cm]    (DaCe)      [below =of Tiramisu]
{\textbf{DaCe (2019)}
: Python embedded DSL \linebreak  
\textbf{Targets:} framework for defining own scheduling primitives based on a data-centric IR called \emph{SDFG}. \textbf{Hwd}: multicore, GPU, FPGA 
\linebreak
\textbf{Scheduling:} affine + user-defined transformations, based on SDFG 
};

\node[SRY, text width=7cm]    (MDH)     [left =of DaCe]
{\textbf{MDH (2023)}
: purely-functional, map-reduce\linebreak  
\textbf{Targets:} correctness checks, autotuning, visualization. \textbf{Hwd}: multicore and GPU
 \linebreak
 \textbf{Scheduling:} single primitive for systematically expressing (de/re)-composition of computations, based on MDH formalism~\cite{10.1145/3665643}

};


\draw[->, very thick] (Tiramisu.south)  to node[below] {} (DaCe.north);
\draw[->, very thick] (Tiramisu.south)  to node[below] {} (MDH.north);
\draw[->, very thick] (Lift.south)  to node[below] {} (MDH.north);

\end{small}
\end{tikzpicture}}\vspace{-2ex}
\caption{Overview of some of the central works focusing on data parallel approaches and how they are related.
}
\label{fig:sched-lang-overview-dp}
\end{center}
\end{figure}
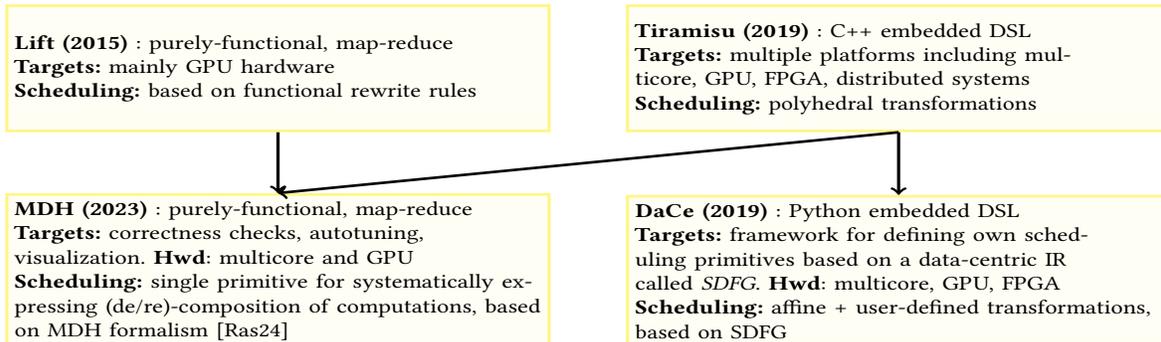

\subsection{Data-Parallel Computations (summarized in \cref{fig:sched-lang-overview-dp})}
\label{subsec:dataparcomp}

\textbf{Lift}~\cite{10.1145/2784731.2784754}, \textbf{Tiramisu}~\cite{8661197}, \textbf{Locus}~\cite{8661203}, \textbf{DaCe}~\cite{10.1145/3295500.3356173}, and \textbf{MDH}~\cite{10.1145/3578360.3580269} address a more general flavor of computations that include image processing and tensor kernels, but also more exotic computations, such as \emph{Probabilistic Record Linkage~(PRL)}~\cite{10.1145/3297280.3297330} from the domain of \emph{Data Mining}, different kinds of \emph{Stencil}~\cite{10.1145/3168824} and climate modeling~\cite{dace-tuning2} applications.\smallskip

\textbf{Lift}~\cite{10.1145/2784731.2784754} (recently also known as \emph{RISE}~\cite{steuwer2022rise}) proposes a restricted purely-functional language, together with a repertoire of code transformations that are expressed as rewrite rules (e.g., map, fusion, fission, stripmining).  Since the optimization space is huge, the rewrites are either directly specified by the user or are guided by procedures that use stochastic search or equality saturation~\cite{koehler2024guided}. More recent work focuses on the design of a strategy language, named {\em ELEVATE}~\cite{10.1145/3408974}, that aims to allow programmers to define new optimization strategies in a composable and reusable way (and is inspired by the work discussed in section~\ref{subsubsec:functional-past}).
Lift mainly targets performance portability across GPU hardware. For example, extending Lift with \texttt{slide} and \texttt{pad} primitives (and associated re-writes) allows efficient computation of individual stencils~\cite{10.1145/3168824} (e.g., by overlapped tiling), and extension with \emph{macro rules} targets matrix multiplication~\cite{10.1145/2884045.2884046}.\smallskip

\textbf{Tiramisu}~\cite{Tiramisu} is a scheduling framework for a C++ embedded domain-specific language (DSL). In contrast to Lift's functional rewrite rules, Tiramisu uses polyhedral transformations, rooted in dependence analysis on arrays.   Similar to Halide, the algorithm specification is separated by the implementation details (hardware, iteration space, and other optimizations).  Unlike Halide, which uses an interval-based representation of iteration spaces, Tiramisu opts for a more mathematically powerful polyhedral representation that supports composition of a complex sequence of affine transformations  
on dense multi-dimensional spaces (i.e., loop nests). 
Notably, Tiramisu uses a four-level intermediate representation (IR) that addresses separation of concerns without unnecessarily barring optimizations: The first level refers to the algorithmic specification, the second to the ordering of computations, the third to the management of data (i.e., storage location on device and layout), and the fourth to the communication between execution nodes. 

Tiramisu supports multiple backends,
including multicore X86 CPUs, Nvidia GPUs (both leveraging LLVM infrastructure), Xilinx FPGAs (Vivado HLS) and distributed machines (MPI), and reports performance results for image stencils, recurrent neural network (dense and sparse), and other non-rectangular iterations spaces that rival those of cuBLAS, cuDNN, Intel OneAPI, and other specialized libraries. 
Recent work~\cite{baghdadi2021deep} reports a learning-based cost model for automatic code optimization that was implemented in the Tiramisu compiler.
%
\smallskip

\textbf{Locus}~\cite{8661203} is a scheduling framework that is aimed at orchestrating the optimization of legacy code written directly in mainstream languages, such as C, C++, Fortran, instead of a (restricted) DSL.  The work addresses the challenges of manipulating large programs written in complex languages, in particular related to expressing clearly and concisely complex collections of transformations---rooted in dependence analysis on arrays---that are applied to (different) code regions, as selected by the programmer.  Locus supports a number of optimization modules off the shelf, as well as procedures for automatically searching the space of code variants.\smallskip

\textbf{DaCe}~\cite{10.1145/3295500.3356173} is a framework that supports a collection of APIs for implementing optimization workflows. {\em The key idea is that the performance engineer uses the APIs to write their own DSLs and optimization pipelines, tailored to the target application}, albeit DaCe comes already equipped with a collection of transformations and optimization passes. 
The most important API is the Stateful DataFlow multiGraph (SDFG), which implements the IR and facilitates the {\em construction} of SDFGs by means of frontends for several language subsets. 
Other APIs refer to transformations on SDFG graphs (e.g., to develop optimization passes and tuners) and code generation APIs, which handle the mapping to different architectures, such as multicore, GPU and FPGA.  

Scheduling optimizations are expressed either on (1) the structural representation, e.g., tiling is represented as nested Map scopes; or, (2) the attributed representation, e.g., ``schedule'' attributes are attached to scope nodes to indicate OpenMP scheduling, GPU thread blocks, etc., and ``storage'' attributes are attached to data nodes to indicate the memory type (CPU heap, GPU global, shared or register memory).
%
For code transformation, DaCe uses a combination of polyhedral and re-write rules on graphs.   
Recent work~\cite{dace-tuning1,dace-tuning2} has reported an autotuning framework that automatically explores the space of some common polyhedral transformations by combining machine learning and performance profiling techniques.\smallskip

\textbf{MDH}~\cite{10.1145/3578360.3580269} is a recent approach that has a particular focus on a structured-language design.
It is grounded in the algebraic formalism of \emph{Multi-Dimensional Homomorphisms (MDH)}~\cite{10.1145/3665643}, and it is aimed to systematically express (de/re)-compositions of computations at each level of the memory and core hierarchies of parallel architectures.  
The claimed advantages are twofold:
{\em First}, MDH catches a multitude of user errors\footnote{
In addition to the errors caught by polyhedral approaches, such as specification of invalid tiling, MDH detects more sophisticated errors, e.g., when the user tries to combine the results of \emph{CUDA Thread Blocks} across invalid memory regions, as per CUDA specification~\cite{cuda-specification}.} and it issues precise error messages for the invalid schedules. 
%
%
{\em Second}, MDH allows leaving any arbitrary optimization decision optionally to its internal autotuning engine (including automatic generation of entire device- and data-optimized schedules), thereby promoting user productivity and performance portability. Similar to Lift and Tiramisu, MDH allows autotuning of straightforward optimization parameters such as tile sizes, but also supports more advanced exploration, e.g., related to layout transformations and memory placement (register/private/shared/global). 
The automatically generated schedules have often been found to offer performance competitive with vendor libraries~\cite{10.1145/3578360.3580269} 
and can also be used as a starting point for manual fine tuning by a performance expert. 
In terms of limitations, MDH's scheduling language is designed to express optimizations at a high abstraction level and is thus unable to express low-level code optimizations (such as loop unrolling, which is handled in MDH internally by heuristics).

\enlargethispage{\baselineskip}

\begin{figure}[H]
\begin{center}
\resizebox{\columnwidth}{1.7in}{
\begin{tikzpicture}[
SRP/.style={rectangle, draw=purple!60, fill=purple!5, very thick, minimum size=1cm},
SRR/.style={rectangle, draw=red!60, fill=red!5, very thick, minimum size=1cm},
]

\begin{small}
\node[SRP, text width=8cm]    (Halide)                              {\textbf{Halide (2013)} -- \linebreak
\textbf{Targets:} fusion of image processing pipelines 
\linebreak
\textbf{Scheduling:} granularity of compute and store,
 \linebreak
split, vectorize, reorder, parallelize};

\node[SRR, text width=7cm]    (TVM)  [right=of Halide]
{\textbf{TVM(2018)} -- 
\linebreak \textbf{Targets:} hardware extensions (tensor cores)
\linebreak         
\textbf{Scheduling:} latency hidding, tensorization,
 \linebreak        
Special Memory Scope};

\node[SRR, text width=7cm]    (Cora)       [below=of Halide] 
{\textbf{CoRa (2022)} -- 
\linebreak \textbf{Targets:} ragged tiles, load balancing
\linebreak
\textbf{Scheduling:} iteration-space splitting,
\linebreak
padding, horizontal fusion, thread remapping};

\node[SRR, text width=7cm]    (Cortex)       [right=of Cora] 
{\textbf{Cortex (2021)} -- 
\linebreak \textbf{Targets:} lowering recursion
\linebreak
\textbf{Scheduling:} graph-level optimizations
};
\end{small}

\draw[->, very thick] (Halide.east)  to node[right] {} (TVM.west);
\draw[->, very thick] (TVM.south)  to node[right] {} (Cora.north);
\draw[->, very thick] (TVM.south)  to node[right] {} (Cortex.north);


\end{tikzpicture}}\vspace{-2ex}
\caption{Overview of some of the central works related to scheduling languages targeting Deep Learning. 
}
\label{fig:sched-lang-overview-dl}
\end{center}
\end{figure}
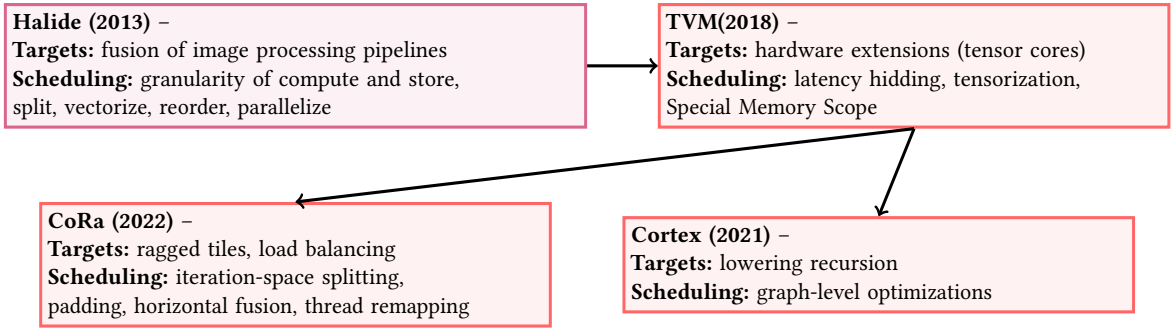

\subsection{Deep Learning Approaches (summarized in \cref{fig:sched-lang-overview-dl})}

Scheduling languages are a common feature in systems focused on code generation for deep learning workloads, supporting computations such as linear algebra, convolutions, and computations for reshaping and splitting data. 
Given the large body of recent work in this area, we focus our discussion here on 
\textbf{TVM}, which 
represents the current state-of-the-art in CPU and GPU mapping of deep learning workloads~\cite{chen2018tvm}.  TVM  achieves performance close to hand-optimized vendor libraries. 
TVM's scheduling language augments Halide schedule constructs with architecture-specific optimizations, e.g., 
applying to TPUs, 
NVIDIA Tensor Cores, and memory hierarchies.
Extensions to TVM's scheduling language were added for \textbf{C\textsc{o}R\textsc{a}}~\cite{CORA}, which supports ragged tiles resulting from problem sizes that are not divisible by the tile size.  These include iteration space splitting, padding, horizontal fusion across operators, and thread remapping for load balance. \textbf{Cortex} provides additional extensions in a dynamic scheduling language to address how recursion is lowered~\cite{Cortex}.
%
Only a limited set of optimizations can be tuned in TVM's scheduling language, and more advanced optimizations require the code generation to be tunable (e.g., based on \emph{Ansor}~\cite{258858} discussed in Section~\ref{subsec:explore-while-autotune}). 
TVM supports \emph{graph-level optimizations} across deep learning operators, particularly fusion of adjacent operators that do not exhibit data dependences. Graph-level optimizations are not accessible from the scheduling language, but are instead performed upon the intermediate representation.   In contrast, operator fusion is implemented within a scheduling language as a loop transformation, as in the SWIRL system for wide SIMD CPUs~\cite{SWIRL}. 
\textbf{Triton}~\cite{Triton}
also targets DNNs. It focuses on block tiling, but does not provide a scheduling language.

\subsection{Graphs (main frameworks summarized below)}
Graph algorithms require a unique set of transformations that reflect structural changes and how to parallelize the execution of graphs.  
\textbf{Galois} 
\cite{irreg-appl-10.1145/1273442.1250759,galois-7185292}  
takes a data-centric view of parallelism where the Parallel program is viewed as Operator + Schedule + Parallel Data Structures.  For Galois, the schedule refers to constructs that relax the order in which graph vertices are processed, later extended with data placement annotations for scalability.  Galois~\cite{nguyen-lenhart-pingali-sched2011} demonstrated that it could serve as a backend for other vertex-centric graph DSLs PowerGraph~\cite{PowerGraph}, GraphChi~\cite{GraphChi} and Ligra~\cite{Ligra}, improving their performance and scalability. 

More recently, \textbf{GraphIt}~\cite{zhang2018graphit}
exposes a scheduling language and a separate 
algorithm language describing operators on vertices and edges. The scheduling language has representations for edge traversal, vertex data layout, and 
structure optimizations. 
The schedules provides various functions to configure edge traversal direction, parallelization, data layout, 
and data placement optimizations. 
It also offers the unique opportunity for structural changes to the graph, such as fusing adjacent vertices into a single vertex.  

\section{Future: Call to Action}
\label{sec:future}

The story portrayed so far shows a past in which performance critical applications were predominantly optimized with general-compiler infrastructure operating on mainstream languages and a present in which we have amassed significant expertise in extracting near-optimal performance for specific computational kernels from various domains --- e.g., image processing, tensor algebra --- by using scheduling languages in conjunction with very restricted programming models (languages) and specialized compiler infrastructure. 

%
However, at its core, the compiler pipeline employed by these systems consists of a combination of well-known code transformations along with architecture-specific optimizations.   A natural next step might 
be to cycle back to the past by gradually lifting the language restrictions, unifying scheduling languages, and  extending their repertoire with more dynamic analyses and support for specifying data layout and movement.   Since writing schedules is challenging, we envision that solutions might: raise the level of abstraction of the scheduling language such that it becomes accessible to domain experts; and/or rely on exploratory compilers that automate schedule generation and selection, while still allowing the user to provide key optimization insights. %
This section predicts several emerging technologies that would benefit scheduling languages in the quest of advancing the holy grail of automating high-performance code generation.


\subsection{Unifying scheduling languages}

%

We have discussed a number of systems, each with its own scheduling language. In several cases, the same code transformation is named differently across systems, e.g., \textit{permute/interchange} and {\textit{flatten/coalesce/collapse}. Conversely, in other cases, very different transformations share the same generic name---see the various kinds of {\em tiling} or the various instances of {\em flattening} irregular nested parallelism~\cite{nesl:vect-models,nesl-on-gpus,cuNESL,data-par-haskell}.

These ambiguities lead to the question: \textit{is standardization of scheduling languages possible, and how can it be achieved?} 
We consider here the alternative strategies for standardization.
First, annotation systems such as OpenMP~\cite{OpenMP1} expose a pragma interface that permits programmers to express thread-level parallelism at a high level, while relying on the underlying compiler to implement the details of the parallelization.  In recent years, OpenMP pragmas describe loop transformations (currently \textit{tile}, and \textit{unroll}).  The strength of standardization is that a committee reviews the language extensions.  However, a weakness is that many of the users of scheduling languages are not part of the high-performance computing community, from which OpenMP arises, and that it takes a long time to add support for a new transformation.

An approach with a potentially broader reach is to integrate the scheduling language into a widely-used open source compiler infrastructure, particularly one designed to support domain-specific systems.  The MLIR compiler~\cite{mlir} provides this capability in the form of the \textit{transform dialect}~\cite{lucke2024mlirtransformdialectcompiler}.  
The transform dialect approach is distinguished from other systems in this paper where a schedule is customized to a particular computation. Instead, the use of the \textit{pdl dialect} to identify applicability of transforms makes the schedules independent of the computation, more like an optional compiler pass, and facilitates “building libraries of composed compiled transforms”~\cite{lucke2024mlirtransformdialectcompiler}.
Therefore, it is not a direct replacement for schedules in other systems, but additional layers on top of transform dialect, such as in PEAK~\cite{PEAK}, could potentially improve this. 
Such an approach is still limited in that it relies on a specific compiler infrastructure that cannot be reused outside that environment.  
Additional transformations may require significant changes to the intermediate representation; e.g., for data layout and data movement optimizations in Section~\ref{subsec:datamovement}.

Alternatively, we can design high-level languages aimed at specifying {\em a (source/target) language together with the code transformations operating on it (them)}.  For example, in the context of functional languages, the essence of a code transformation can often be concisely and elegantly expressed in the form of inference rules, declared for each syntactic category of the source language. We hypothesize that the verbose, complex and error-prone code that implements the code transformation---i.e., that traverses the abstract-syntax tree, matches the pre-conditions and applies the inference rules---can be automatically generated.
Rather than requiring all compiler experts to work on the same code base, such an approach would democratize the implementation of a new language and its optimizing compiler by allowing to reuse, adjust and compose available components from the specification of the code transformations belonging to other languages. Such solutions can build on the findings of strategy languages, such as {\it ELEVATE}~\cite{10.1145/3408974} and MDH~\cite{10.1145/3578360.3580269}, which are aimed to improve the language aspects of schedules: clearly-defined semantics, composibility and opportunity for (automatic) verification. For example, it is arguably easier to reason about and verify a high-level specification rather than its low-level implementation.  

%


Standardization of scheduling languages could make all of these approaches viable.  A commitment by the community towards standardization is the first step.

\subsection{Data layout and data movement as part of the schedule}
\label{subsec:datamovement}

Since data movement is the key cost in terms of time and energy, incorporating specifications of data layout and data movement into scheduling languages should be a universal part of future exploratory compilers.  
Organizing data according to its access pattern then requires modifying the computation accordingly.  We discussed the role of data layout in terms of sparse tensor computations, but other computations also benefit from data layout and data movement specifications.  

Historically, data copy was applied to reorganize submatrices, especially to avoid conflict misses in cache or stage data in explicitly managed storage~\cite{Temam93,Saday08}. CHiLL incorporated datacopy into its scheduling language~\cite{Chen2007CHiLLA}, which was later adapted in CUDA-CHiLL to copy data to/from global memory, shared memory, and texture memory in GPUs~\cite{CUDA-CHiLL}.  More recently, Fireiron and MDH enrich these data movement specifications for GPUs~\cite{hagedorn2020fireiron,10.1145/3578360.3580269}.
Follow-on work, for example in Graphene, optimizes data layout and thread mapping, particularly to prepare data and computation for tensor cores~\cite{Graphene}.  

Other layouts beyond strided rectangular regions and sparse tensor representations have been shown to reduce data movement.  For example, fine-grain data blocking, where logically adjacent three-dimensional subdomains are stored in contiguous memory, have been shown to significantly reduce data movement in structured grid computations~\cite{bricks}.  A two-dimensional \textit{tile}, also a fine-grain data block, has been used in Triton to accelerate deep learning computations~\cite{Triton}.

Looking forward, data layout specifications in scheduling languages should support these and other future layouts.  To generalize this approach, the scheduling language should be integrated into a compiler supporting high-level abstraction such as MLIR.  To generate code, the compiler must map between \textit{logical} and \textit{physical} data layouts, so that address calculations can map logical indices to their physical locations.  In sparse tensor computations, where logical indices may not have corresponding physical entries, the inverse mapping from physical to logical indices can be used to find corresponding elements in other tensors during co-iteration, as is done in \texttt{dlcomp}~\cite{dlcomp}.  The mapping from a variety of layouts to hierarchical sublayouts could be supported with additional logical dimensions,
generalizing the approach in Graphene~\cite{Graphene}.


\subsection{Architecture-specific scheduling primitives}
Related to data layout is the need to support architecture-specific hardware features, which was discussed above for Graphene~\cite{Graphene} and Fireiron~\cite{hagedorn2020fireiron} in the context of matrix processing units for NVIDIA GPUs.  How might architecture-specific constructs be integrated into a scheduling language?

One mechanism is to layer scheduling languages, providing an extensible programming language approach to build new scheduling constructs either from existing ones, or by describing how to rewrite code based on scheduling constructs.
More recently, Exocompilation~\cite{ikarashi2022exocompilation} addresses this challenge with a user-level scheduling language --a Python-embedded DSL called Exo -- that permits a performance expert to describe architecture details that impact performance.  In Exo, users may define specialized memories including lower precision data types; semantics of hardware instructions, e.g., fused multiply-add; and, hardware state so as to avoid redundant computation or data movement in composing operations.  

The above approaches address the extensibility of scheduling languages, but not the widespread adoption of these extensions; i.e., unification and raising the level of abstraction.  As new classes of architectures arise, extensions permit exploration of how to derive performant schedules.  Once the community settles on a solution, these architecture-specific layers can be integrated into the language and compiler mechanisms.

\subsection{Integration with runtime}

Many optimizations require a combination of static and runtime information to make decisions.  For example, inspector-executor strategies, as previously discussed in the context of sparse tensors, may wish to choose data layouts based on the nonzero structure of the sparse tensors -- information only available at runtime. Similarly, a runtime data layout transformation, such as ZMorton order,  
would need to be part of the schedule, especially since it may require costly layout conversion that needs to happen early in the computation.   

Furthermore, in computations such as preconditioners for sparse solvers, the resulting loop parallelism 
is data-dependent \cite{venkat:sc16}, hence the computation needs to be 
reorganized in a sequence of parallel wavefronts, whose structure is determined at runtime. 
In molecular-dynamics (astrophysics) simulations, the distribution of molecules (particles) is unknown at least until runtime, and it might also dynamically change. 
Dynamic analyses~\cite{ data-iter-reorder-1,data-iter-reorder-2} proposed in these contexts use a statically-generated inspector to perform data- and (loop-) iteration reordering in order to optimize spatial and temporal locality, respectively. Such runtime-reordering transformations have been later integrated in sparse-polyhedral frameworks~\cite{STROUT201632} and have further inspired the notion of locality hypergraphs~\cite{data-iter-reorder-2}, for example, as a way to model the optimization of communication in distributed computations as graph-partitioning problems~\cite{temp-hypergraph-mmm,sudoku}.

Finally, another facet of runtime integration is the decision process for selecting the appropriate optimized code variant if that choice is data dependent.  The selection might involve a lookup into pre-autotuned variants based on data ranges, or use of inference based on problem size or other problem size features.

Current scheduling languages are typically weak at supporting integration with the runtime system, of the kind described above. We hypothesize that future research directions will be aimed at incorporating such dynamic analyses into the repertoire supported by scheduling languages.

\subsection{Raising the level of abstraction}

Most of the work reviewed so far expresses a schedule as a (precise and detailed) sequencing of compiler transformations.  Arguably, such a representation is unfriendly to domain experts, who might be well versed in (or at least willing to learn) ``parallel thinking'', but may find that the gap to ``thinking like a compiler'' is too wide to bridge. Another disadvantage is that it requires (even) the compiler experts to run the optimization recipe in their head and to specify in detail its effects on the entire code base --- which might be impractical for large computations.  Ideally, the appropriate level of the scheduling language should specify
{\em concepts that are accessible to domain experts}, 
and from which {\em a sequence of compiler transformations is automatically inferred}.  

A first possible approach could be to pair up the original program with a specification of the high-level (loop) structure of the target (optimized) code. For example, the original program could be annotated with (comment) labels identifying loops and code fragments and the labeled names can be used in the specification to declare the target code structure.   The challenge resides in {\em inferring the sequence of (loop) code transformations whose application conforms with the target code structure} and in designing a clear semantics of the specification language that is intuitive and promotes ease of use.
Since code transformations are guaranteed to be correct, the resulted program is guaranteed to be semantically equivalent to the original one, but the question of when the desired structure is met needs to be answered in an unambiguous way that also matches user's intuition.

\begin{figure}
\begin{subfigure}{0.43\columnwidth}
  \includegraphics[width=0.75\columnwidth]{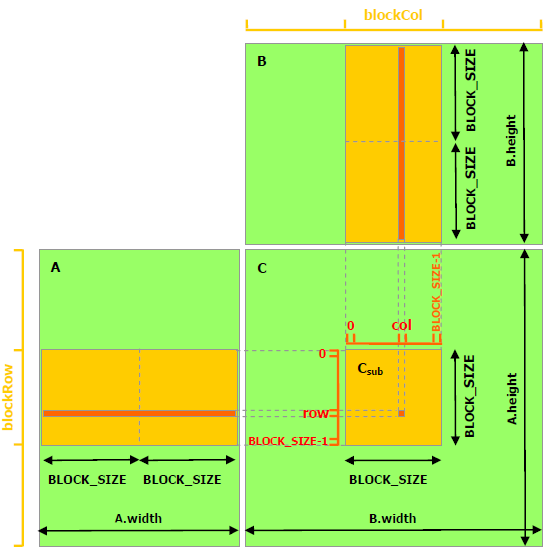}
\caption{Picture schedule for Matrix Multiplication (MM)}
\label{fig:picture-mm}
\end{subfigure}
\begin{subfigure}{0.55\columnwidth}
  \includegraphics[width=0.99\columnwidth]{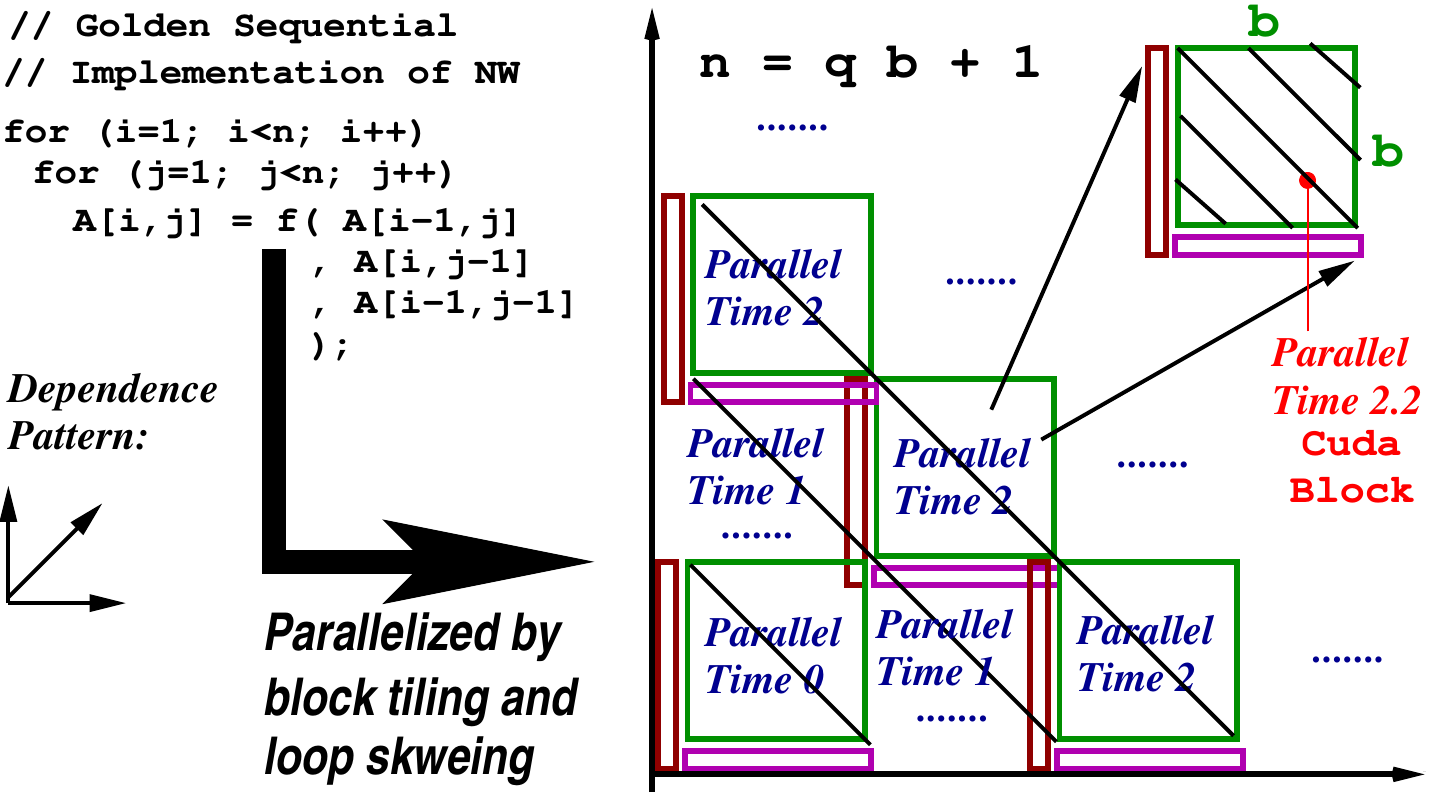}
\caption{Picture Schedule for Needleman-Wunsch (NW).}
\label{fig:picture-nw}
\end{subfigure}\vspace{-2ex}
\caption{Picture-like Schedules for MM~\cite{kirkhwu} and NW~\cite{sc22mem}: they seem isomorphic with a language representing array slices, i.e., describing the result (and input) slices produce (read) at each level of the parallel-hierarchy of the hardware.}
\label{fig:picture-sched}
\end{figure}

The second possible approach is to describe by means of pictures the array slices produced/used at each level of the hardware, respectively.
Figure~\ref{fig:picture-sched} illustrates pictures used in two scientific publications to explain locality optimizations for matrix multiplication~\cite{kirkhwu} and for the Needleman-Wunsch problem~\cite{sc22mem}. The former requires tiling at each hardware level, and the latter requires tiling and two loop-skewing transformations.    While the domain experts might find it difficult to ``think like a compiler" --- e.g., loop skewing has seldom been called an intuitive transformation --- they could arguably reason in terms of wavefront parallelism, both across the blocks and within a $b\times b$ block depicted in figure~\ref{fig:picture-nw}.\footnote{
Figure~\ref{fig:picture-nw} hints that the computation is organized as a wavefront, where the $i^{th}$ anti-diagonal of $b\times b$ blocks is processed at parallel time $i$. Each $b\times b$ logical block is processed within a Cuda block of size $b$: the green box denotes the write set of a Cuda block and the red/purple vertical/horizontal lines denote the read set of a Cuda block; all are being mapped to shared memory. The computation inside a Cuda block is also organized in a wavefront structure that computes at parallel time $i$ the $i^{th}$ anti-diagonal (see the zoomed block).
} In a nutshell, the domain expert could possibly draw a picture that describes how the result is produced at each level of the hardware (e.g., grid, block, and thread level for GPUs). A first step in this direction is taken by MDH~\cite{10.1145/3578360.3580269}.  

Such a picture-representation of schedules is essentially isomorphic with a language of array slices, i.e.,  which specifies what slice of the result is produced at each hardware level {\em and} from what input slices.  Similar as before, the challenging step refers to automatically inferring the sequence of code transformations from the picture (slice language), and verifying that it actually produces the desired picture.

\section{Related Work}
\label{sec:relwork}

This section surveys several strands of related work, as follows: Section~\ref{subsec:heuristic-explore-before-tuning} surveys approaches that refer to the classical, heuristic-based compilation and, as well, to compiler-exploration techniques that produce a multitude of code variants, which are then autotuned to the particularities of hardware and datasets. Section~\ref{subsec:explore-while-autotune} surveys approaches that deeply integrate the autotuning in the compiler-exploration process by testing and sampling code variants during compilation.  Finally, section~\ref{subsec:autotuning-and-ML} examines advanced autotuning solutions, in particular the ones based on machine-learning (ML) algorithms.

\subsection{Heuristic Compilation and Compiler Exploration}
\label{subsec:heuristic-explore-before-tuning}

While scheduling-based compilers are becoming increasingly attractive in the HPC community, there is also a wide range of promising compiler approaches that do not expose optimization to the user in the form of scheduling programs, or expose optimizations at all.

\subsubsection{Classical Compiler Approaches} 
Classical compilation approaches automatically generate low-level parallel code according to a set of internal (optimization) heuristics. As such, they do not necessitate user intervention, but may be somewhat customizable by means of pragma annotations or compiler flags. Examples include tensor DSLs such as TC~\cite{vasilache2018tensor,10.1145/3355606} and FlexTensor~\cite{zheng2020flextensor}, and polyhedral~\cite{PolyhedralOpt} compilers such as Pluto, PPCG~\cite{PolyPluto1,verdoolaege2013polyhedral,verdoolaege2017scheduling}. 
They have demonstrated that efficient parallelization of (at least) affine programs is possible for multi-core and GPU hardware.  
In turn, such analyses are facilitated by the development of analytic cost models for locality of reference, capable of deriving, for example, asymptotic lower bounds at the level of misses in a set-associative cache hierarchy~\cite{AnModelCache}.
As well, since the affine domain is too restrictive for many practical applications, various techniques were devised to utilize explicit (user) annotations within polyhedral transformations of otherwise unanalyzable patterns. For example, Pencil allows the user to summarize read-write access sets of unanalyzable functions~\cite{pencil}, and other works~\cite{pencil-red,poly-with-annot} allows to integrate reduction, loop- and task-parallel OpenMP annotations within the polyhedral optimizer. 
%
%
Such heuristic-based approaches can be considered more productive than scheduling-based compilers, because the user is not in charge of making complex optimization decisions. However, they are fundamentally limited by the fact that they essentially produce {\em one} optimized code version: Even when this code version is parameterized by tile sizes and the like, it still cannot cover the case when different sequences of code transformations are needed to optimize different classes of datasets, i.e., ``one size does not fit all''.

\subsubsection{Compiler Exploration Followed By Autotuning}  More recent approaches provide means to adapt the compilation strategy to the hardware and dataset characteristics. A class of compilers, such as CLBlast~\cite{10.1145/3204919.3204924} and MDH~\cite{10.1145/3665643}\footnote{MDH also exposes optimizations in the form of schedules, as discussed in Section~\ref{subsec:dataparcomp}}, do not rely on schedules, but instead they expose their optimization spaces in the form of (formally defined) parameter spaces, which further enables automatic exploration of such spaces by means of autotuning frameworks~\cite{10.1145/2628071.2628092,7328205,VANWERKHOVEN2019347,PETROVIC2020161,wu2023ytopt,hellsten2023baco,10.1145/3427093}.   However, a parameter-based space makes it difficult for humans to express optimizations, because parameters lack the language properties that are specifically designed for human interaction.

Another instance of compiler exploration that does not rely on a scheduling language is proposed by Futhark: a purely-functional language that supports nested compositions of the full set of data-parallel operators---such as map, reduce(-by-index), scan, scatter.
Here, the (arbitrarily-nested) application parallelism is mapped to the hardware by an exploratory procedure, dubbed ``incremental flattening''~\cite{futhark-incflat}, that utilizes map fission and map-loop interchange to create semantically-equivalent code versions that systematically map more and more levels of application parallelism to the hierarchy supported by the hardware. The code versions are independently optimized 
and combined together into a program by branching on predicates that compare a dynamic program measure (e.g., the degree of utilized parallelism) with a threshold. The best combination of code versions is derived by autotuning the threshold values: this is implemented by a deterministic procedure that is guaranteed to produce the optimal result in minimal number of runs, as long as the dynamic measure conforms with a monotonic property~\cite{futhark-autotuning}. However, tile sizes and the like do not conform with this monotonic property and are not autotuned. Of note, the use of a scheduling language is not practical in Futhark, because the language supports reverse-mode automatic differentiation~\cite{futhark-ad-sc22}: this is a complex program-level transformation that produces code that cannot be ``predicted'', hence ``scheduled'', by the user.

A variety of image processing DSLs, compilation techniques and autotuners have been developed independent of Halide and PolyMage. Modesto~\cite{modesto} (CPU+GPU), Absinthe~\cite{Absinthe} (CPU) and Stencil-Gen~\cite{Stencil-Gen} (GPU) rely on analytical models to determine the optimal schedule among a multitude of variants generated using various tiling strategies (including overlapped tiling with streaming), storage optimizations and (greedy) fusion heuristics.   Another body of work refers to applying dynamic analysis to optimize stencils: OPS~\cite{ops} uses a combination of delayed execution and dependence analysis to resolve at runtime hindrances to static analysis that typically occur in large applications, and ARTEMIS~\cite{artemis} uses bottleneck analysis via runtime profiling to guide the application of optimizations, and the tuning of various code generation parameters.  Other work explores (1) scalable and adaptive autotuning frameworks for stencil computations~\cite{csTuner,StencilMART,GSTuner} and (2) efficient mapping of image-processing pipelines to FPGA hardware~\cite{stencil-fpga-1-polymage,stencil-fpga2}

In comparison with approaches based on scheduling languages, the drawbacks of the frameworks surveyed in this section are possibly (1) suboptimal performance, since there is no guarantee that the optimization space is exhaustively explored; and, (2) high compilation time, because code variants need to be explicitly executed on the target system during autotuning.
On the plus side, such frameworks may offer more complete languages that can express more general flavors of computations, for example, enabling efficient execution of applications from domains such as finance~\cite{pricing2}, remote-sensing~\cite{bfast,stl-time-series} and computer vision~\cite{approx-nn}.

\enlargethispage{\baselineskip}

\subsection{Integrating Scheduling Languages with Autotuning}
\label{subsec:explore-while-autotune}

Systems that use scheduling languages employ different techniques for deriving 
the optimized sequence of transformations described by the scheduling language. 
Here, we use the phrase \textit{defining the optimization search space} to refer to
the process of how possible optimization sequences are determined and specified.  A \textit{point in the search space} is then an instantiated schedule that can be provided to the system to generate code.  
These search spaces are typically prohibitively large, and it is impractical to evaluate each point in the search space.  Therefore, various techniques are needed to limit the search space and sample a subset of points, which we refer to as \textit{search space pruning}.  Instead of the performance models used in compiler heuristics as discussed in the previous subsection, autotuning
is frequently used to measure the execution time of an optimized code variant running on a specific hardware platform.   As previously discussed, autotuning systems arose because of the increasing challenge of deriving accurate performance models as architectural complexity has exploded.

A very common approach to defining the optimization search space 
is to use manually-written scheduling templates, written by performance experts, with variables embedded in each template that are placeholders for parameters in the search space; e.g., in frameworks for deep learning such as AutoTVM~\cite{chen2018learning} and SWIRL~\cite{SWIRL}. In domain-specific systems, where the set of possible optimizations is well understood, this approach can be effective because the search space can be kept narrow. 

In other systems, the search space is generated by an algorithmic procedure, using various techniques to limit the size of the search space.  CUDA-CHiLL~\cite{CUDA-CHiLL} avoids combinatoric growth in the search space by separating the exploration of data placement in the GPU memory hierarchy (global/shared/texture memories or registers) from tuning optimization parameters (thread, block, tile and unroll factor sizes), treating them as independent variables.  
Protuner~\cite{haj2020protuner} uses Monte Carlo Tree Search to look ahead when generating and evaluating complete schedules. FlexTensor~\cite{zheng2020flextensor} generates schedules by enumerating different combinations, based on a specific ordering within the schedule space.  Ansor \cite{258858} uses evolutionary search and learned cost model to find optimized TVM schedules. This approach is superior to the approach of AutoTVM since it extends to optimizing operators not available as templates and captures complex optimization patterns. 

For sparse tensor computations, Ahrens et al.~\cite{ahrens2022autoscheduling} co-optimizes the computation and format of a sparse tensor.  This system outputs a schedule by ranking and filtering by asymptotic complexities and runtime, which reduces the scheduling space by orders of magnitude and generate kernels which perform asymptotically better than the default TACO schedules.
WACO~\cite{won2023waco} co-optimization  considers both sparse formats and sparse schedules together. A sparse CNN model extracts the feature set from sparse programs, and a superschedule explores the combined search space using the Approximate Nearest Neighbour Search (ANNS).

\subsection{Tuning using Machine Learning (ML)}
\label{subsec:autotuning-and-ML}

An alternative approach to conventional autotuning is to construct a predictive model using deep learning: a learning phase derives an associated cost with different schedules and inputs, and inference performs a  model lookup at runtime.   While a search is still needed to build the model, an accurate model can be reused, and leverages the large investment in deep learning systems to make this efficient.  Early work by Park et al. developed a predictive model for iterative compilation, called a tournament predictor~\cite{Park11}.

{ImageCL} \cite{falch-elster-ml-2015, falch-elster-cpe.4384}, introduced in 2015, showed how to use Machine Learning(ML) for autotuning CUDA codes targeting stencils for image processing.  It was later extended as a more general framework {AUMA}\cite{falch-elster-ml-2015, auma-falch-elster-cpe.4029,falch-elster-cpe.4384}, for OpenCL. ImageCL and AUMA are DSLs whose schedules are autogenerated based on the model derived by ML training. 
Branches from AUMA's opensource repository~\cite{AUMA} include extensions for Julia as well as 3D extensions for Adaptive Mesh Refinement.

Like with conventional autotuning, the challenges with this approach revolve around navigating a prohibitively large search space during the training phase.  For instance,~\cite{baghdadi2021deep} claims to have spent three weeks on training data collection.
This is even after restricting the number of schedules per program to 32.
Nonetheless, once trained, ML-based optimization techniques continue to exhibit state-of-the-art performance and some prominent work in this domain are~\cite{Park11,kulkarni2012mitigating,halide19, haj2020neurovectorizer, cummins2017b, zheng2020flextensor, chenAutoTVM}. 

Patabandi et al. identify a key challenge in the cost of training to be a \textit{Multiplicative Domain Formulation (MDF)}~\cite{Patabandi:CC23}.  This term refers to the size of search spaces for multiple transformations through a combinatorial exploration of their schedule domains.  For convolutions, they demonstrate an example of \textit{Additive Domain Formulation (ADF)}, querying an existing model for one transformation~\cite{ml4copt} when building the model for another transformation, demonstrating a 100$\times$ reduction in inference time, while maintaining high accuracy. 

\section{Summary}
\label{sec:conclusions}


This paper has provided a taxonomy of scheduling languages that illustrates how the past approaches aimed at optimizing performance-critical applications have motivated the emergence of scheduling languages, and how future improvements may cycle back to the past. 

Past work ($1997-2012$) has targeted scientific applications expressed in mainstream languages, e.g., in terms of loop nests, that could not be adequately optimized by the heuristic pipelines of general-purpose compilers.  This has led to autotuning libraries and to more advanced compilation techniques that better explore the optimization space. Examples include iterative compilation that, at each step, generates multiple candidates and selects the one that performs best on the target hardware; or, rewrite-rule systems that encode algorithmic properties of high-level language constructs. 

Present work ($2013-2023$) is driven by the observation that specialization gives rise to performance, in that restricting (and purifying) the specification language (DSLs) allows it to be effectively optimized by means of a relatively small number of code transformations, which are sequenced either explicitly by the compiler expert, or automatically by tuning strategies.  

We envision that future work will broaden the specification language to cover more general computations and the repertoire of code transformations available to scheduling, e.g., with various dynamic analyses and with support for specifying data layout and data movement. On the one hand this will require improvements to the search strategies such as to cover this expanded space in practical time. On the other hand, this will require productivity-oriented improvements that minimize the user interaction with the scheduling language, while still allowing specification of key optimization insight. Ideally, the domain expert can directly interact with scheduling languages, by lifting their level of abstraction.   Achieving this requires a level of integration and automation reminiscent of earlier exploratory compilers, thus circling back to the past.

\paragraph{Acknowledgments.}
We thank the Lorentz Center and its 2022 Workshop on Generic Auto-tuning for GPU Applications, which brought the authors together, and Tor Andre Haugdahl for his contributions re. Tiramisu.

This work was partially funded by the \emph{Deutsche Forschungsgemeinschaft (DFG, German Research Foundation)}~--~project~\emph{PPP-DL}~(470527619), the US National Science Foundation project CCF-2107556, and SFI-CGF through Research Council of Norway Project no. 309960. 
\bibliographystyle{alpha}
\bibliography{sample,references}

\end{document}

%% file: textResources/lang-listing.tex
\lstdefinelanguage{specLang}
{
  morekeywords={
    do,
    else,
    float,
    for,
    fun,
    if,
    in,
    int,
    inout,
    include,
    let,
    loop,
    struct,
    then,
    type,
    val,
    void,
    while,
    with,
    module,
    where,
    sort,
    multired,
    task,
    instance,
    name,
    task,
    subtask,
    variant,
    run_at,
    calls,
    tunable,
    distribution,
    rchop,
    mappar,
    mapreduce
  },
  sensitive=true, 
  morecomment=[l]{//}, 
  morecomment=[s]{\{-}{-\}}, 
  morestring=[b]", 
  literate={\\}{\fn}{1} {->}{$\rightarrow$}{1} {<-}{$\leftarrow$}{1},
}

\definecolor{eclipseBlue}{RGB}{42,0.0,255}
\definecolor{eclipseGreen}{RGB}{63,127,95}
\definecolor{eclipsePurple}{RGB}{127,0,85}